\documentclass[10pt,journal,compsoc]{IEEEtran}

\usepackage{cite}
\usepackage{epstopdf}
\usepackage{amsmath}
\usepackage{mathtools}
\usepackage{mathrsfs}
\usepackage{subfig}
\usepackage{graphicx}
\usepackage{multirow}
\usepackage{epsfig}
\usepackage{verbatim}
\usepackage{color}
\usepackage[english]{babel}
\usepackage{array}
\usepackage{makecell}
\usepackage{epsfig}
\usepackage{multirow}

\usepackage[linesnumbered,ruled]{algorithm2e}
\begin{document}
%
\title{Discriminative Learning-based \\Smartphone Indoor Localization}

\author{Jos\'e Luis Carrera V.,~\IEEEmembership{Member,~IEEE,}
        Zhongliang Zhao,~\IEEEmembership{Member,~IEEE,}
        Torsten Braun,~\IEEEmembership{Member,~IEEE}
        Haiyong Luo,~\IEEEmembership{Member,~IEEE,}
        Fang Zhao~\IEEEmembership{}
\IEEEcompsocitemizethanks{\IEEEcompsocthanksitem J. Carrera, Z. Zhao, and T. Braun are with the Institute of Computer Science, University of Bern, Switzerland; H. Luo is with the Institute of Computing Technology, Chinese Academy of Science, China; F. Zhao is with the School of Software Engineering, Beijing University of Post and Telecommunications, China. \protect
}
\IEEEcompsocitemizethanks{\IEEEcompsocthanksitem Corresponding author: Z. Zhao (zhao@inf.unibe.ch)\protect\\
}
}

\markboth{Submitted to IEEE Transactions on Mobile Computing}%
{Shell \MakeLowercase{\textit{et al.}}: Discriminative Learning-based Smartphone Indoor Localization}

\IEEEtitleabstractindextext{%
\begin{abstract}
Due to the growing area of ubiquitous mobile applications, indoor localization of smartphones has become an interesting research topic. Most of the current indoor localization systems rely on intensive site survey to achieve high accuracy. In this work, we propose an efficient smartphones indoor localization system that is able to reduce the site survey effort while still achieving high localization accuracy. Our system is built by fusing a variety of signals, such as Wi-Fi received signal strength indicator, magnetic field and floor plan information in an enhanced particle filter. To achieve high and stable performance, we first apply discriminative learning models to integrate Wi-Fi and magnetic field readings to achieve room level landmark detection. Further, we integrate landmark detection, range-based localization models, with a graph-based discretized system state representation. Because our approach requires only discriminative learning-based room level landmark detections, the time spent in the learning phase is significantly reduced compared to traditional Wi-Fi fingerprinting or landmark-based approaches. We conduct experimental studies to evaluate our system in an office-like indoor environment. Experiment results show that our system can significantly reduce the learning efforts, and the localization method can achieve performance with an average localization error of 1.55 meters.
\end{abstract}

}

\maketitle

\IEEEdisplaynontitleabstractindextext

%
\IEEEpeerreviewmaketitle

\ifCLASSOPTIONcompsoc
\IEEEraisesectionheading{\section{Introduction}\label{sec:introduction}}
\else
\section{Introduction}
\label{sec:introduction}
\fi

%
%
%
%
\IEEEPARstart{N}owadays, location information has become a transcendent part of the daily life of mobile users. The growing area of ubiquitous computing technology has increased the demand of location-based services. The main goal of indoor localization is to supply to a mobile device a mechanism to locate itself and provide this information to users. However, contrary to Global Positioning System (GPS) for outdoor environments, currently there is not a simple and accurate solution for indoor localization. Typically, indoor position-based services require higher localization accuracy than outdoor services. Additionally, the often limited computation and power resources on mobile devices put strong constraints on the algorithmic complexity design of solutions. Thus, indoor localization is still considered an open challenging problem.

With the availability of more embedded sensors on mobile devices, numerous indoor localization techniques (i.e, ultrasonic, radio frequency tags (RFID), magnetic field, Wi-Fi, Bluetooth, etc.) have been proposed in recent years. However, due to the wide availability of Wi-Fi signals in indoor environments, Wi-Fi radio-based localization has attracted most attention among different localization approaches. Wi-Fi radio-based localization relies on measured radio parameters, such as signal power to estimate the absolute positioning of targets \cite{ARealTimeIndoorTracking}. Wi-Fi received signal strength indicator (RSSI) is the most widely used parameter for indoor localization.

Radio-based indoor localization can be classified as range-based and range-free methods. Range is defined as the propagation distance from the target to Anchor Nodes (ANs). Fingerprinting is a range-free approach, which is often used because of its robustness to multi-path propagation. In Wi-Fi fingerprinting techniques, a training or learning phase is needed to build the RSSI signature database off-line. In the positioning or on-line phase, the target location is estimated according to the results of comparing RSSI readings and the signature database. 

Indoor environments provide many different ambient radio signals, such as Wi-Fi, Bluetooth, magnetic field, sound, etc \cite{SemanticSLAM}. The set of observed ambient radio signals provides unique radio signal signatures for many points in indoor environments. These radio signal signatures can be used as landmarks to detect unique locations in the environment. Thus, the radio signal signatures are often used as the fingerprint of the related location. However, building the fingerprinting database of radio signals is a labor-intensive and time-consuming task. Moreover, the fingerprint database is prone to environment changes. This means whenever there is a modification of the physical indoor environment, the previous database will be outdated, and a new one has to be built. Therefore, the traditional fingerprinting approach is not suitable for large-scale scenarios. Some localization methods have been proposed considering magnetic field measurements. By using some advanced data fusion mechanisms (i.e., particle filters), some works propose to use magnetic field readings as alternative \cite{ReliabilityAugmented} or as complement to Wi-Fi signatures \cite{Magicol} to further improve the positioning accuracy. However, due to the intrinsic disturbances caused by electrical appliances and building materials, the change of the magnetic field with location is quite significant in indoor environments. Moreover, magnetic field fingerprints can not be unique, which yields to localization errors.

Indoor localization can be seen as a likelihood estimation problem. Thereby, locating a target is the problem of estimating the location of a target given a set of available observations \cite{unscentedPF}. To achieve high accuracy, Kalman filters or particle filters are often used to fuse this set of observations to derive accurate localization estimations. However, filter-based data fusion normally requires intensive computation resources, which makes the real-time localization a challenging task for smartphones with limited resources. Therefore, optimization of the data fusion algorithm is key for efficient smartphone indoor localization systems.

In this work, we propose a discriminative learning-based smartphone indoor localization system aiming at reducing the site survey efforts while attaining high localization accuracy. Our approach provides localization by fusing Wi-Fi radio readings, magnetic field readings and floor plan information in an enhanced particle filter. Our approach is able to exploit historical environmental information from a learning phase as well as on-line information to provide high and stable localization accuracy. Our approach integrates discriminative learning-based landmark detection, range-based localization models, and a graph-based floor plan representation in an enhanced particle filter. Moreover, since our localization method exploits landmark detection with room level accuracy, we reduce the effort and time in building the learning phase compared to traditional fingerprinting-based and landmark-based localization approaches. Our landmark fingerprint database is built by taking Wi-Fi and magnetic field measurements while walking randomly through the environment, which requires only room labeled samples in a very short time period.

We prototype our approach on commodity smartphones, with all the localization algorithms run on the smartphone itself. To validate our localization system, we conduct extensive experiments in a complex office-like environment. Evaluation results show that our approach can achieve an average localization error of $1.55m$ with standard deviation of $0.73m$ with significantly reduced site survey efforts, which makes it suitable for large-scale deployments. The main contributions of this work are as follows.
\begin{itemize}

 \item We propose an efficient indoor localization system for smartphones that is able to significantly reduce the site survey effort while still achieving high localization accuracy. Since our approach requires only room level landmark detection (i.e., room recognition), the landmark fingerprint database is built by collecting Wi-Fi and magnetic field measurements by simply walking randomly through the environment. Thus, we reduce cost, manpower, effort, complexity and survey time in building the learning phase compared to traditional fingerprinting-based and landmark-based localization approaches. 
 
	\item We propose an enhanced particle filter to integrate machine learning-based landmark detections, range-based localization models and a graph-based floor plan representation to provide highly accurate and stable localization results. Our approach is able to fuse historical environmental information from a learning phase and perform on-line localization with stable localization accuracy.

    \item We perform a set of experiments and analysis to validate the performance of our localization method. Several configuration parameters, such as the number of particles as well as the number and position of Wi-Fi access points are tested to show that our system can significantly reduce the site survey efforts while preserving high accuracy.

\end{itemize}
The rest of the paper is organized as follows. In Section II we present some related work. The localization approach is reviewed in Section III. Section IV presents the implementation of the terminal-based system and the experiments' setup. Section V presents the performance evaluation results of our approach. Section VI concludes the paper.

\section{Related Work}
The current commercial and social importance of indoor location-based applications has attracted significant attention in recent years. Indeed, indoor localization has been investigated lately and many solutions have been presented. Radio frequency (RF) based approaches include technologies such as GPS, wireless local area networks (WLAN) and RFID localization. Non-RF-based solutions include laser,  audio, visual, ultrasonic and infrared  sensors \cite{neurocomputing}. In \cite{Multi-camera}, a non-RF-based localization approach is proposed. This solution estimates the user location based on surveillance cameras. Any wearable device is needed by the user. However, the calibration process for the system becomes complex. In \cite{ActiveOffice} the localization system is deployed using special infrastructures such as infrared and acoustic sensors. Despite the fact that the localization system can achieve high accuracy, building these kinds of infrastructure is very costly. Due to the high vulnerability to  environmental disruptions, most of the non-RF-based localization approaches require demanding calibration processes \cite{neurocomputing}. 

Among RF-based indoor localization approaches, trilateration has been regarded as a popular solution \cite{Trilateration1} \cite{Trilateration2}. Trilateration determines the position of the target based on the distances to some anchor nodes. However, unlike outdoor localization, trilateration does not work well in indoor environments because of the presence of obstacles and room partitions \cite{Wi-FiFingerprinting}. Therefore, as an alternative to trilateration, the process of signal collection and association with indoor locations has become a promising approach for indoor environments. This process is called fingerprinting. Fingerprinting approaches are conducted in two phases: the learning phase is performed off-line and the query phase that is executed on-line. The off-line phase is conducted to build or update a \textit{$<$signature, location$>$} database, which consists of a set of reference points with known coordinates and the signatures collected from available sensors. Then, the on-line phase is aimed to find the closest match between the features of the measured signatures and those stored in the database. \cite{RealTimeIndoorNavigation}. RADAR \cite{RADAR} was the first work that utilized Wi-Fi fingerprinting. In RADAR, the interest area is divided into a grid of 1x1 m. RSS measurements are taken at each cell intersection to create the radio map database. Then in the on-line phase, RSS received is compared to the radio map database to estimate the target location.

In addition to RF signals, the earth's magnetic field $MF_{geo}$ and magnetic field fluctuations in indoor environments can also be potentially leveraged for indoor localization. There are several works that show the feasibility of using anomalies of the magnetic field to provide indoor localization \cite{ReliabilityAugmented}\cite{Magicol}\cite{IndooMobileRobot}\cite{SignalSLAM}. However, due to  ferromagnetic material and electrical objects, magnetic signatures have many ambiguities in indoor environments (i.e., similar magnetic field value in different locations). Therefore, MF measurements should be fused with other indoor sensor measurements to derive accurate locations.

Recently, landmark-based approaches have been proposed. Similar to fingerprinting-based approaches, landmark-based approaches utilize smartphone sensors to detect unique positions in the indoor environment \cite{SemanticSLAM} \cite{FusingWiFiLandmark}. In \cite{SemanticSLAM}, two types of landmarks are defined: seed landmarks and organic landmarks. Seed landmarks are positions that can be associated with specific objects in the environment such as stairs and elevators and are used to bootstrap the system. Organic landmarks are detected based on their unique signature on the sensors, and are identified by ambient radio signatures such as magnetic field or Wi-Fi. The defined landmarks are used to calibrate the localization errors, which means that a carefully-designed landmark selection procedure is the key to high performance. Although, landmark-based localization approaches can achieve high accuracy, the localization process relies heavily on the presence of numerous predefined landmarks in the environment (e.g., stairs, escalators, elevators, doors). Thus, building the landmark fingerprint database becomes an intensive effort demanding process.

Various machine learning-based approaches have been proposed that uses fingerprinting to estimate user indoor locations. Machine learning-based indoor localization can be classified into generative or discriminative methods, which builds the model using a joint probability or conditional probability respectively \cite{FingerprintComparison} \cite{Generative+Discriminative}. K-nearest neighbors (KNN) is the most basic and popular discriminative technique. Based on a similarity measure such as a distance function, the KNN algorithm determines the $k$ closest matches in the signal space to the target. Then, the location of the target can be estimated as the average of the coordinates of the $k$ neighbors \cite{KNN1}. Generative localization methods apply statistical approaches, e.g., Hidden Markov Model \cite{HiddenMarkov}, Bayesian Inference \cite{BayesianInference},  Gaussian Processes \cite{GaussianSignal}, on the fingerprint database. Thus, the accuracy can obviously be improved by adding more measurements. In \cite{GaussianSignal} for instance, Gaussian Processes are used to estimate the signal propagation model through an indoor environment. There is a limited number of works that have focused in reducing off-line efforts in learning-based approaches for indoor localization \cite{24} \cite{25} \cite{26}. These approaches reduce the off-line effort by reducing  either the number of samples collected at each survey point or the number of survey points or both of them (i.e., reducing number of collected samples and number of survey points). Then, a generative model is applied to reinforce the sample collection data. In \cite{24} for instance, a linear interpolation method is used. In \cite{25}, a Bayesian model is applied. In \cite{26}, authors propose an propagation method to generate data from collected samples. In \cite{Generative+Discriminative}, authors combine characteristics of generative and discriminative models in a hybrid model. Although this hybrid model reduces offline efforts, it still relies on a number of samples collected from fixed survey points (i.e., labeled samples) along the environment. Therefore, to maintain high accuracy, the number of survey points shall be increased in larger environments. Thus, collecting samples from numerous survey points will become a demanding process, which makes the system unsuitable to large environments.

Although generative approaches can handle the missing value problem, discriminative approaches often achieve better performance. Generally speaking, all the machine learning-based localization methods using fingerprinting can achieve good accuracy, if a large number of labeled samples are available. However, such a sample collection process could take severals hours or days for small or big areas, which is very labor expensive and time consuming. Therefore, it is essential to reduce the efforts in offline sample collection procedures while still maintaining high localization accuracy.   

Therefore, this work proposes to combine radio signal transmission features along with magnetic field disturbance to reduce ambiguity conflicts. Moreover, our approach incorporates a room level landmark detection in the localization process to accurately converge to the actual position. Different from previous landmark-based approaches, we need only few  landmarks, which significantly reduces site survey efforts. We only define subareas in the environment as landmarks (e.g. rooms). Then, the target is first located at room level, and then applying our ranging model and particle filter algorithm, its  location inside the room is derived.

\section{System Overview}
\begin{figure}
	\centering
	\includegraphics[scale=0.33]{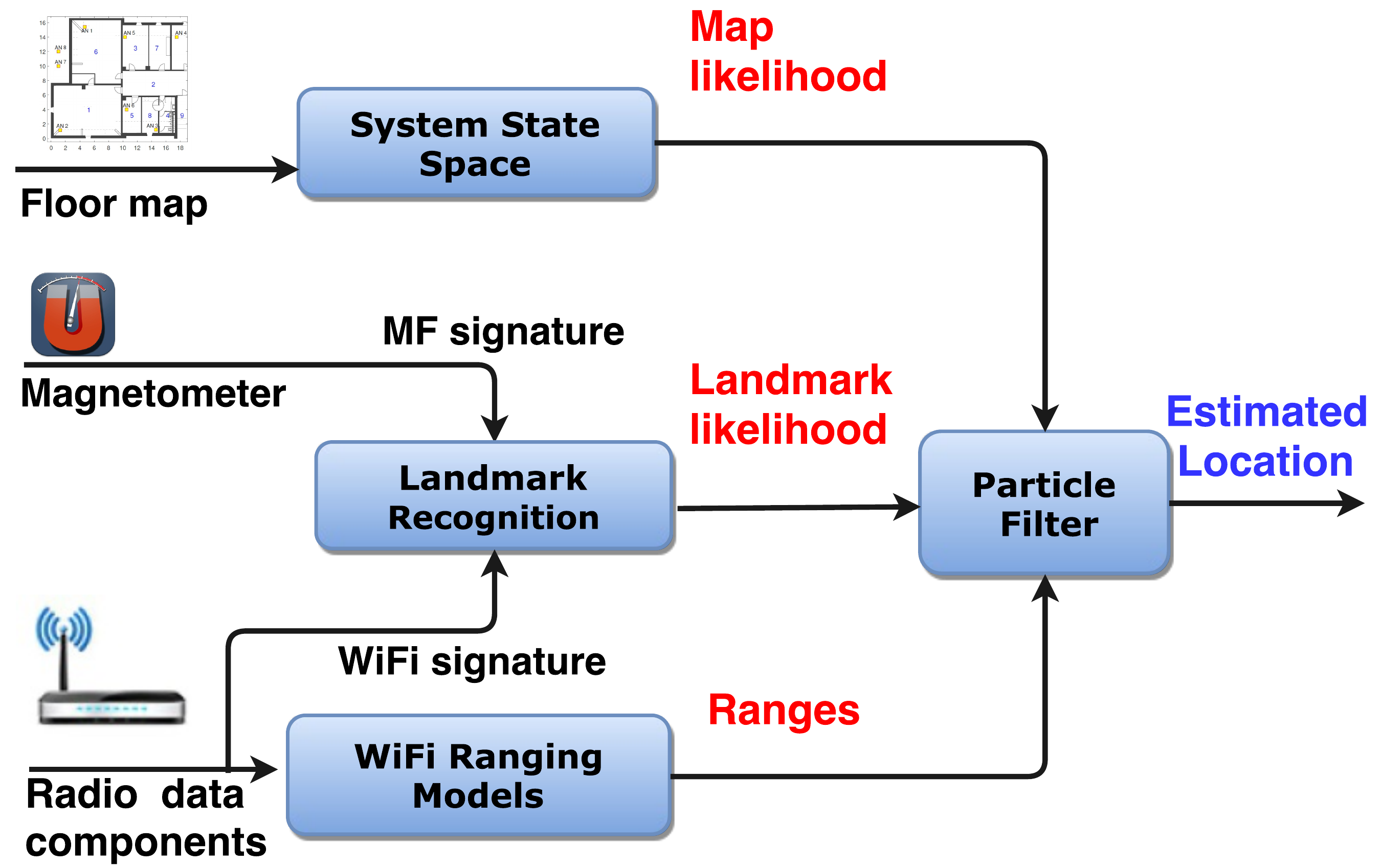}
	\caption{Indoor Localization System Architecture.}
	\label{systemArchitecture}
\end{figure}

This section presents the design details of the proposed indoor localization problem. Figure \ref{systemArchitecture} summarizes the structure of our proposed approach, which includes four key subcomponents: a discrete system state space model, a discriminative learning-based room landmark recognition module, a Wi-Fi ranging model, and an enhanced particle filer module. Details of each subcomponent are described below.

\subsection{Discrete System State Space}
\begin{figure}
	\centering
	\includegraphics[scale=0.22]{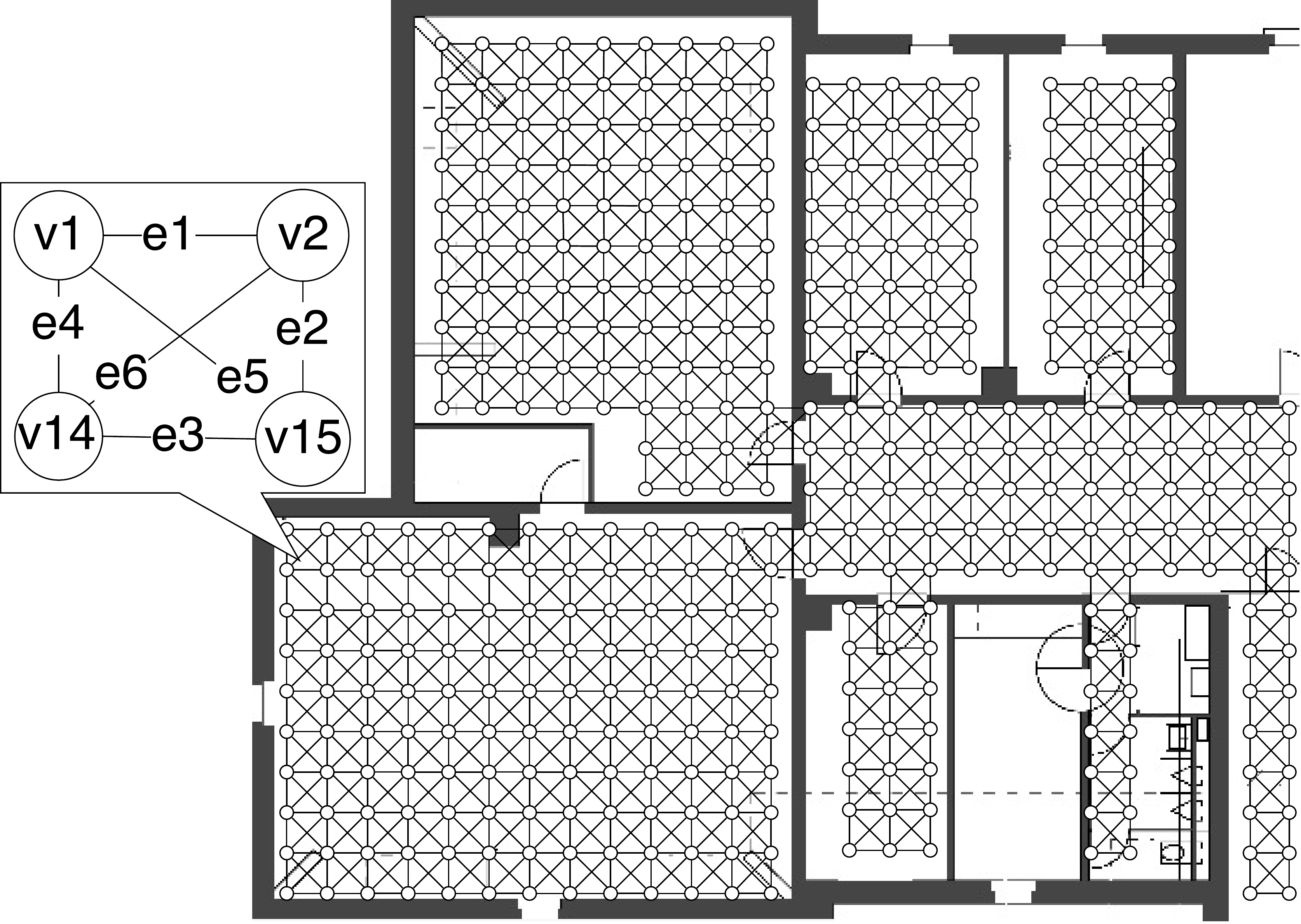}
	\caption{Discrete System  State Space. Nodes (circles) are interconnected by edges. Edges define the transition model nodes.}
	\label{DiscreteSpace}
\end{figure}

In order to minimize computational overhead, our system defines a discrete structure to replace the conventional floor map. Thus, all the system states (i.e., indoor positions) can be represented by a discrete set of locations by converting from a continuous state space to a discrete state space. Figure \ref{DiscreteSpace} shows an example of a graphical representation of the discrete state space of our system. Nodes represent positions in the physical environment, whereas edges define the transition model between nodes. For instance, in Figure \ref{DiscreteSpace} $e1$ defines a potential transition from node $v1$ to node  $v2$ and vise-versa. Thus, the essential connectivity and accessibility of a complex indoor environment is represented as an undirected graph. We consider the physical environment $G$ as a spatial data structure that defines space as an array of nodes arranged in rows and columns. $G$ can be defined as follows:
\begin{equation}
 G=(\Upsilon,I,\xi),
 \end{equation}
where $\Upsilon$ is a set of nodes. $I$ is a set of features defining each node $\upsilon$ in $\Upsilon$. $\xi$ is a set of edges $e$ defining connections between nodes $\upsilon$. The grid extraction can be carried out in a few simple steps as follows:
\begin{itemize}
	\item Specifying the areas where it is permissible to walk in rooms, corridors, etc.
    \item Building a grid of nodes for each permitted area. Nodes are separated by $0.25m$ from each other in the room and column. Each node corresponds to an element $\upsilon$ in $\Upsilon$.  
\item Generating the subset of features $i \in I$ for each $\upsilon \in \Upsilon$. Each subset $i$ contains three elements related to a $\upsilon$ element: coordinates $(X ,Y)$ and an identifier of the room to which $\upsilon$ belongs. 
\item Generating set $\xi$ of node connections. Each node is connected with their immediate  neighbors in the grid.
\end{itemize}

\subsection{Discriminative Learning-based Landmark Detection}
\label{landmarDetection}
We propose to use a discriminative learning-based approach to provide landmark recognition with room level accuracy. For the purpose of indoor localization, we consider a landmark as a location point where readings of at least one sensor show a stable and discriminative pattern. This pattern is referred as the signature of the landmark. We define each area separated by walls as an independent room landmark in the indoor environment. In this manner, the process of building the landmark fingerprint database is simplified to collect some fingerprints by walking randomly through each predefined room. Thereby, the required effort  and time  to accomplish this task is significantly reduced compared to traditional fingerprinting-based and landmark-based localization approaches. 

Most of the indoor environments offer some ubiquitous radio signatures in different sensing domains \cite{SemanticSLAM}. In this work, we propose to use Wi-Fi radio and  magnetic field signatures to provide room landmark detection. Thus, landmark fingerprints are composed by Wi-Fi and magnetic field signatures.  The tuple $<fingerprint, location>$ defines the instances in the landmark fingerprint database.

Wi-Fi-based landmark detection is one of the most widely used techniques, since it can make use of already deployed Wi-Fi access points (APs) or anchor nodes (ANs). Additionally, it can provide landmark detection without any knowledge of the access point location or the signal propagation model \cite{RealTimeIndoorNavigation}. However, physical configuration of the indoor environment can influence accuracy. Unlike most of the existing landmark-based approaches that define many landmarks to correct the localization error, we use landmarks for a different purpose. In our approach, we only define landmarks at room level, and the landmark detection is applied to recognize which room a target is. Afterwards, we use this information and apply our ranging model and particle filter algorithm to further derive more accurate locations within the detected room. Therefore, we do not need to measure a lot of landmarks, which is a labor intensive task. 

Indoor magnetic field data combine the geomagnetic field $MF_{geo}$ of the Earth and the magnetic field from ferromagnetic objects $MF_{env}$. Magnetic field readings change significantly with locations, but the magnetic field is stable over time \cite{ReliabilityAugmented}. However, distortions of the magnetic field produced by ferromagnetic material and electrical objects lead to  ambiguities in the magnetic field in indoor environments. Thus, using magnetic field fingerprints alone is not accurate.\cite{IndooMobileRobot}. 

A smartphone's magnetometer measures the indoor magnetic field vector $MF_{p}$. Equation \ref{EMagneticFiled} shows the relationship among $MF_{p}$, $MF_{geo}$ and $MF_{env}$ in the absence of noise.

\begin{equation}\label{EMagneticFiled}
	MF_{p}=MF_{geo}+MF_{env},
\end{equation}  

where $MF_{p}=(MF_{x},MF_{y},MF_{z})$ is a three-dimensional vector representing the magnetic field value on the phone local coordinate system. Therefore, $MF_{p}$ becomes error prone due to its phone orientation dependence. As an alternative and similar to \cite{ReliabilityAugmented}, we propose to use the gravity sensor to extract the vertical $MF_v$ and horizontal $MF_h$ component of the magnetic field. Thus, the magnetic field signature is defined by the vector $MF=(MF_v,MF_h)$.

According to the No Free Lunch theorem \cite{NoFreeLunch}, there is no single learning algorithm that universally performs best across all application domains. Therefore, for indoor localization problems, a number of classifiers should be tested to validate their performance under a specific indoor environment. In fingerprinting-based landmark detection approaches, the k-nearest Neighbors (k-NN) method is a widely used technique. However, k-NN provides poor estimation performance due to the large variability of radio signals \cite{neurocomputing}. An alternative approach is to consider the localization problem as an approximation function problem. Therefore, discriminative learning models such as KStar, Multilayer Perceptron (MLP), J48 and Support Vector Machine (SVM) can be applied for fingerprinting localization purposes. 

We setup the room landmark detection method based on some experimental result analysis. Experiments were performed with KStar, Multilayer Perceptron (MLP), J48 and Support Vector Machine (SVM). Results show that in our specific setups, KStar overcomes MLP, J48, SVM in terms of landmark detection accuracy. Moreover, KStar is able to provide high room landmark detection accuracy.  Thereby, we set up KStar as discriminative algorithm to provide room landmark recognition in our localization approach. KStar is an instance-based learner. It means that classification is performed based on pre-classified samples. The fundamental assumption is that similar instances should have similar classifications \cite{KStar}. In KStar, similarity is established by a distance function. Thus, distance between instances is defined as the complexity of transforming one instance into another.  A detailed evaluation of the proposed landmark detection model and performance comparison of different machine learning algorithms can be found in subsection \ref{subsection:result_landmark}.

\subsection{Ranging Method} \label{rangeEstimation}
With the derived room information, we focus on how to use the ranging model to derive more accurate location within that room. Range is defined as the propagation distance between the target and the ANs. Ranges can be derived by using signal parameters such as RSSI. In theory, RSSI monotonically decreases with increasing propagation distance \cite{PassiveWi-FiFineGrained}. Some models to relate RSSI to the propagation distance have been proposed e.g., Log Distance Path Loss (LDPL) \cite{PathLossModel}, Nonlinear Regression Model (NLR) \cite{PassiveWi-FiFineGrained}. These models are called radio propagation models. 

The radio propagation model that describes the LDPL can be described by Equation \ref{LDPLEquation}: 
\begin{equation}\label{LDPLEquation}
	P(r)=P(r_0)-10 \cdot \gamma \cdot \log_{10}(\frac{r}{r_0}),
\end{equation}  
where $P(r)$ is the received signal power in a certain location, $P(r_0)$ refers to the power loss in a free space. $\gamma$ is the path loss efficient. Therefore, $r$ can be calculated as follows:
\begin{equation}
	r=10^{(\frac{P(r_0)-P(r)}{10\cdot\gamma})},
\end{equation}  

The NLR model is defined by Equation \ref{NLREquation}:  
\begin{equation}\label{NLREquation}
	r=\alpha \cdot e^{(\beta \cdot P(r))},
\end{equation}  
where $d$ is the distance between the transmitter and the receiver. Both $\alpha$ and $\beta$ are environmental variables.

In indoor environments the accuracy of the radio propagation models is affected by multipath and Non Line of Sight (NLOS) propagation \cite{PassiveWi-FiFineGrained}.  Thus, to reduce ranging errors introduced by NLOS and multipath propagation, we propose a new propagation model by combining LDPL and NLR. 

In LDPL, $P(r_0)$ is defined under Line of Sight (LOS) propagation (i.e., free space), whereas, NLR parameters are estimated by considering LOS and NLOS propagation.  Thereby, it is expected that LDPL will perform better under LOS and short distances, whereas NLR will achieve better performance under NLOS and large distances. Hence, based on these observations, we define our ranging model as follows:

\begin{equation}
 r=
 \begin{cases}
    10^{(\frac{P(r_0)-P(r)}{10\cdot\gamma})}       & \quad \text{if } \text{  P(r) $> P(r_{5})$ dB}\\
    \alpha \cdot e^{(\beta \cdot P(r))}  & \quad \text{if } P(r) \text{ $\leq P(r_{5})$ dB}\\
  \end{cases}
\label{eqRangingModel}
\end{equation}  
where $P(r_{5})$ is the received signal power under LOS conditions (i.e., in short distance, usually $5m$).

\subsection{Enhanced Particle Filter}
Filtering is the problem of estimating the state of a system when a set of observations becomes available on-line \cite{unscentedPF}. We consider indoor localization as a filtering problem, in which the convergence results predict the position of the target from a set of environmental measurements. Hence, the key component of our localization approach is the enhanced particle filter we propose. Particle filter is a powerful tool that is able to represent a probability distribution function (PDF) over the target area that represents the indoor environment \cite{unscentedPF}. Indoor localization can be modeled as estimating system state by processing a sequence of noisy measurements. Particle filters estimate the posterior distribution of the state system based on some conditioned measurements $Z_t$ at time $t$ \cite{ARealTimeIndoorTracking}.  Thus, we define the system state vector $X_t$ at time \textit{t} as follows:

\begin{equation}
	X_{t}=[x_t,y_t],
\end{equation}  
where $(x_t,y_t)$ are the Cartesian coordinates of the target object. The set of \textit{N} weighted particles can be defined as:
\begin{equation}
	P_{t}=[X_{t}^{i},W_{t}^{i}], i=1,...,N,
\end{equation}  
where$X_{t}^{i}$ is the state vector with $W_{t}^{i}$ of the \textit{i}th particle at time \textit{t}.

\begin{algorithm}

\SetKwInOut{Input}{Input}
\SetKwInOut{Output}{Output}

\Input{$P_{t-1}=[X_{t-1}^i,W_{t-1}^i]$,State Space, $N^{'}\%$}
\Output{$P_{t}=[X_{t}^i,W_{t}^i]$}
Order $X_{t-1}^i$ ascending according $w_{t-1}^i$\\ 
Sampling step\\
\ForEach{n=1,...$N^{'} \cdot N/100$ }
 {
  Position $x_{t-1}$= draw uniformly distributed 
   }
\ForEach{n=1,...$N^{'} \cdot N/100$,...$N$ }
 {
  Position $x_{t-1}$= randomly chosen from immediate neighbor from $\Upsilon$
 }
Update $P_{t}$ from $P_{t-1}$\\
\ForEach{$X_{t-1}$}
 {
  Set $x_{t}$=$x_{t-1}$ 
 }
\caption{Sampling Function}
\label{samplingFunctio}
\end{algorithm}

\subsubsection{Particle Initialization and Sampling}
Sampling function decides the location of particles. Algorithm \ref{samplingFunctio} describes the sampling function \textit{S}. At the beginning of the localization process, $N$ particles are generated uniformly distributed over the whole system state space. Equal weight $W_{t}^{i}=\frac{1}{N}, i \in [1,N]$ is assigned to each particle in $P_{t}$. After initialization, all particles move according to the sampling function \textit{S}. Function \textit{S} introduces variety based on $P_{t}$ as follows:

\begin{itemize}
	\item $N^{'}\%$ of the particles are spread uniformly distributed over the whole System State Space $G$. $N^{'}$ is the percentage of particles with lower weights. $N^{'}$ is an empirically chosen constant.
    \item The position of particles with higher weights is updated based on a random function. There are two possibilities, either the particle stays on the same position $v \in \Upsilon$ or it is updated to one immediate neighbor on $\Upsilon$.
\end{itemize}

\subsubsection{Weight Update and Resampling}
The associated weights $W_t^i$ of the propagated particles must be corrected after updating their positions. The associated weight should be updated based on the likelihood of the observations conditioned on each particle $p(Z_t\mid X_t^i)$ at time $t$. The observation vector is defined by the estimated ranges to different ANs and the result of the machine learning-based room landmark detection approach. Thus, the range observation vector at time $t$ is defined as $Zr_{t}=[d_t^j], j=1...N_{AN}$, where $N_{AN}$ is the number of ANs. The room landmark observation vector at time $t$ is defined as $Zld_{t}$. We can assume that the ranging and the landmark detection process are independent from each other. Thus, the probability $p(Z_t\mid X_t^i)$ can be determined as follows:

\begin{equation}
\begin{split}
	p(Z_t\mid X_t^i) & =p(Zr_t\mid X_t^i) \cdot p(Zld_t\mid X_t^i) \\
   &=p(d_t^j\mid X_t^i) \cdot p(Zld_t\mid X_t^i)
\end{split}    
\end{equation}

In this phase, the associated weight $W_t^i$ of each particle  is given by the ranging and room landmark information. There, at time $t$, small weight values will be assigned to particles with low probability to observe $Zr_t$ and $Zld_t$ in their positions. Particles with large associated weights will have a stronger contribution in the determination of the state belief of the system.  
In order to avoid confusion between different likelihoods used in this work, hereafter we refer to $p(d_t\mid X_t^i)$ as the ranging likelihood, $p(Zld_t \mid X_t^i)$ as the room landmark likelihood and $p(Z_t \mid X_t^i)$ as the overall likelihood.

We can assume that the ranges to different ANs are independent from each other. Therefore, the ranging likelihood can be defined as follows:
\begin{equation}
p(Zr_t\mid X_t^i)=\prod_{j=1}^{N}p(\hat{d}_{j,t} \mid X_t^i),
\end{equation}

where $\hat{d}_{j,t}$ is the measured distance to the AN $j$ at time $t$. Hereafter, $p(\hat{d}_{j,t} \mid X_t^i)$ will be referred as the individual likelihood. Each individual likelihood can be written as:

\begin{equation}
p(\hat{d}_{j,t}\mid X_t^i)=\frac{1}{\sigma_j \sqrt{2\pi}}\exp^{- \frac{[\hat{d}_{j,t}-\sqrt{(x^i-x_j)^2+(y^i-y_j)^2}]^2}{2\sigma_j^2}},
\end{equation}
where $(x_j,y_j)$ are the coordinates of the $j$th AN.

In complex indoor environments Wi-Fi signals suffer from some random variations during transmission. This random behaviour is produced by the presence of multiple obstacles such as walls, furniture, ceiling, etc. Obstacles introduce a mixed transmission between Line of Sight (LOS) and Non-LOS (NLOS) conditions. NLOS propagation introduces a significant bias in power-based ranging \cite{PhaseU}. To mitigate the influence of ranging errors on the definition of the ranging likelihood $p(\hat{d_k\mid X_t^i})$, we propose to adopt the same weighting technique used in our previous work \cite{IndoorTrackingFusing}. The weighting technique magnifies the contribution of the individual likelihood with smaller errors and suppresses the contribution of larger ranging errors. Therefore, the weighting technique is defined on each individual likelihood as follows:
\begin{equation}
p(Z_t \mid X_t^i)=\prod_{j=1}^Np(\hat{d}_j \mid X_t^i)^{m_j},
\end{equation}
where $m_j$ is the exponential weight for the individual likelihood of the $j$th AN. In general, estimation of larger distances introduce more errors than small distances. Thus, the exponential weight $m_j$ can be defined as being inversely proportional to the estimated range outputs \cite{IndoorTrackingFusing}: 
\begin{equation}
m_j=\frac{\frac{1}{d_j}}{\sum_{n=1}^{N_{ap}} \frac{1}{d_n}},
\end{equation}
where $N_{ap}$ is the number of ANs.

Room landmark likelihood $p(Zld_t\mid X_t^i)$ is provided by the machine learning-based room landmark detection approach. In section \ref{landmarDetection} we analyze the performance of some machine learning algorithms to detect landmarks in our indoor environment. These experiments show that the KStar algorithm achieves the best landmark detection performance. Thus, KStar is selected as machine learning landmark detection algorithm for our proposed localization approach. The KStar \cite{KStar} classifier calculates the entropy-based distance sums over all possible transformations between two instances. Then, the probability function is defined as the probability of all transformational paths from instance $a$ to $b$.

The resampling step is a crucial but computationally expensive component of a particle filter approach. This phase is adopted to eliminate particles with small associated weights by replacing them by  particles with large associated weights. Thus, after updating each particle in the prediction phase, we perform a resampling process in a systematic manner. The resampling process in this phase relies on the individual likelihood $p(d_t\mid X_t^i)$ and the room landmark likelihood $p(Zld_t \mid X_t^i)$. It means that the associated weight of each particle is calculated by using observations related to range and room landmark estimations. Afterwards, the weighted center given by all the particles is calculated as the estimated position of the target object.

\section{Implementation and Experiments}
\subsection{Implementation details}

\begin{figure}
	\centering
	\includegraphics[scale=0.14]{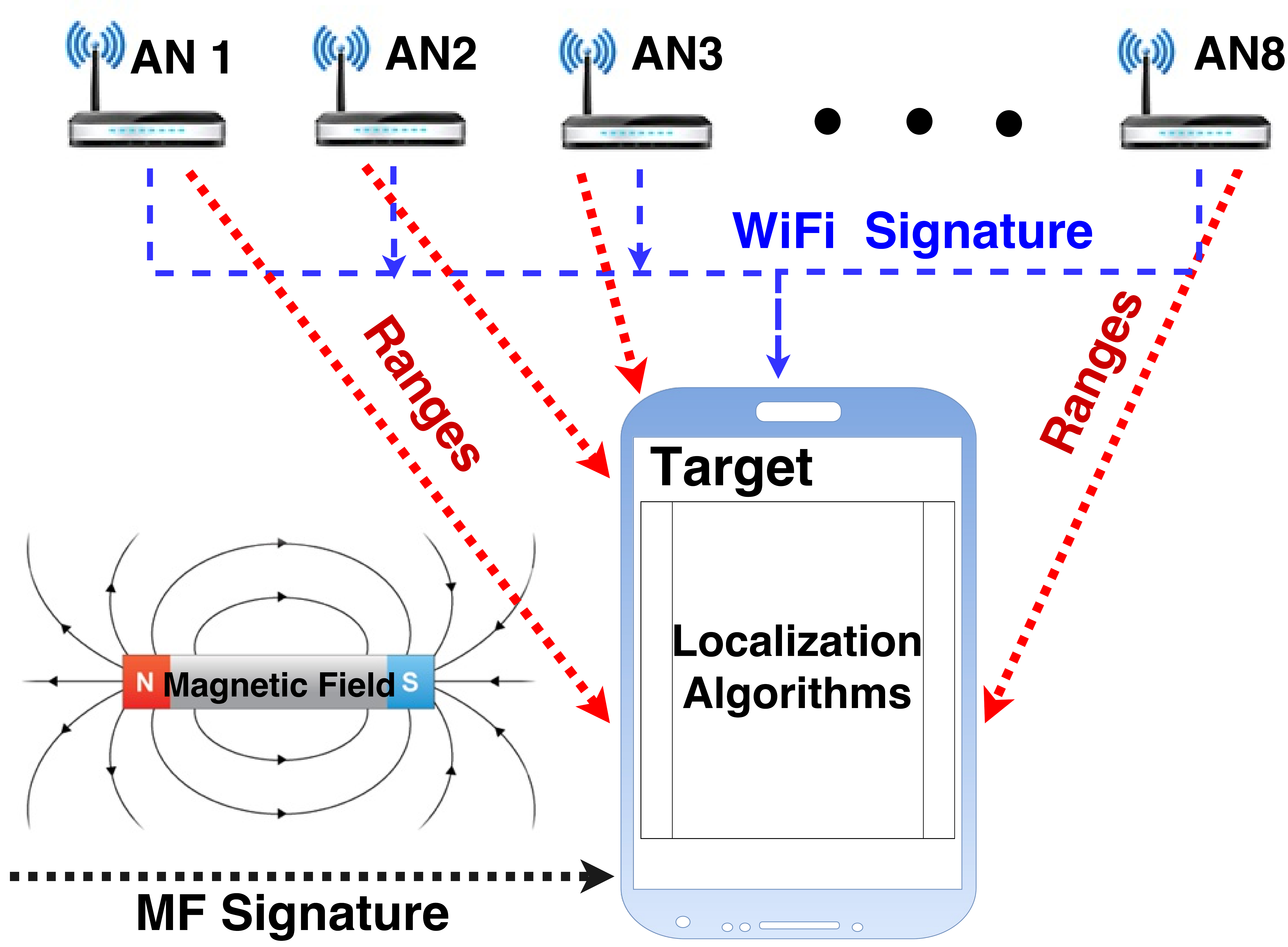}
	\caption{System Implementation Overview. The localization algorithms are deployed in a smartphone.}
	\label{SystemOverview}
\end{figure}

We have implemented a terminal-based system for accurate indoor localization. The system comprises two main components: a mobile target and many Wi-Fi Anchor Nodes (ANs). The proposed localization algorithms are running on the mobile target. Figure \ref{SystemOverview} presents the overview of the system.

ANs are some commercial Wi-Fi access points deployed at known or unknown locations along the area of interest. Positions of the known location ANs are chosen to provide the maximum coverage inside the area of interest. Thus, the locations of ANs are defined on the boundary corners and the boundary itself. We have adopted D-Link D-635 and D-Link DAP-2553 as Wi-Fi ANs in this work. The beacon period is configured to $100ms$ for ANs. 

The mobile target can be any commercial Android smartphone, which supports Wi-Fi RSSI readings and magnetic field sensors readings. We have implemented the localization algorithms in a Motorola Nexus 6 smartphone. Hereafter, we refer to  Motorola Nexus 6 as mobile target (MT). In order to save resources in the smartphone, we set the sampling rate of magnetic field sensor to 14Hz. However, the Wi-Fi sampling frequency is much lower, i.e., 3Hz. Table \ref{table_smarphone} shows the main characteristics of the mobile target used in this work. 

Additionally, it is necessary to know the floor plan of the area of interest. The system requires information related with restricted areas such as sizes of walls and corridors. Particles are not allowed to be spread over restricted areas. The system reports the location of the target in real time.

\begin{table}
	\centering
	\caption{Mobile Target Specifications}
	\begin{tabular}{|l|l|} \hline
	\textbf{Model}&\textbf{Platform}\\ \hline
	\multirow{7}{*}{MT }
	&\textbf{Model:}  Motorola Nexus 6; \textbf{OS:} Android 5.1.1\\
	&\textbf{CPU:} Quad-core 2.7 GHz; \textbf{RAM:} 3GB\\
	&\textbf{WLAN:}Wi-Fi a/b/g/n\\
	&\textbf{Accelerometer:}\\
	&Resolution:0.039 Range:19.613\\
	&\textbf{Magnetometer:}\\
	&Resolution: 0.150 Range:9830\\
	   \hline\end{tabular}
	\label{table_smarphone}
\end{table}

We have conducted a set of experiments in office-like indoor environments to evaluate the performance of our proposed localization system. Experiments were conducted in the building of the Institute of Computer Science (INF) at the University of Bern. A part of the third floor with an area of $288 m^2$ ($18 m$ x $16 m$) was chosen to deploy the  localization system. To evaluate the performance of each subcomponent in the proposed architecture, we perform individual experiments to validate the performance of the discriminative learning-based room landmark detection module and different signal propagation models in the ranging module respectively.

\subsection{Room Landmark Detection Experiment Setups}
\begin{figure*}[htp]
	\centering
	\subfloat[Scenario 1 (5 ANs)]{\includegraphics[scale=0.29]{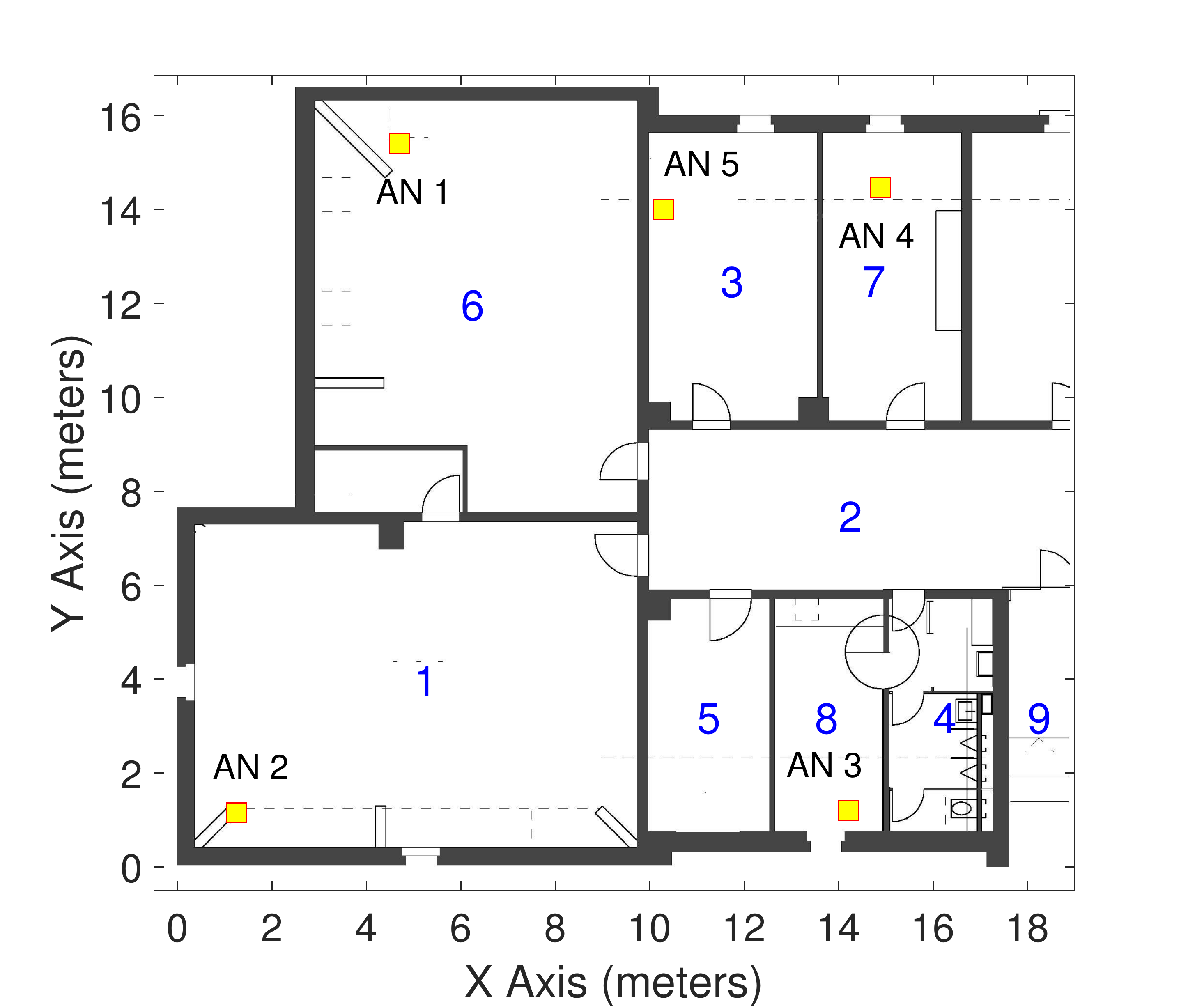}}
	\subfloat[Scenario 2 (8 ANs)]{\includegraphics[scale=0.275]{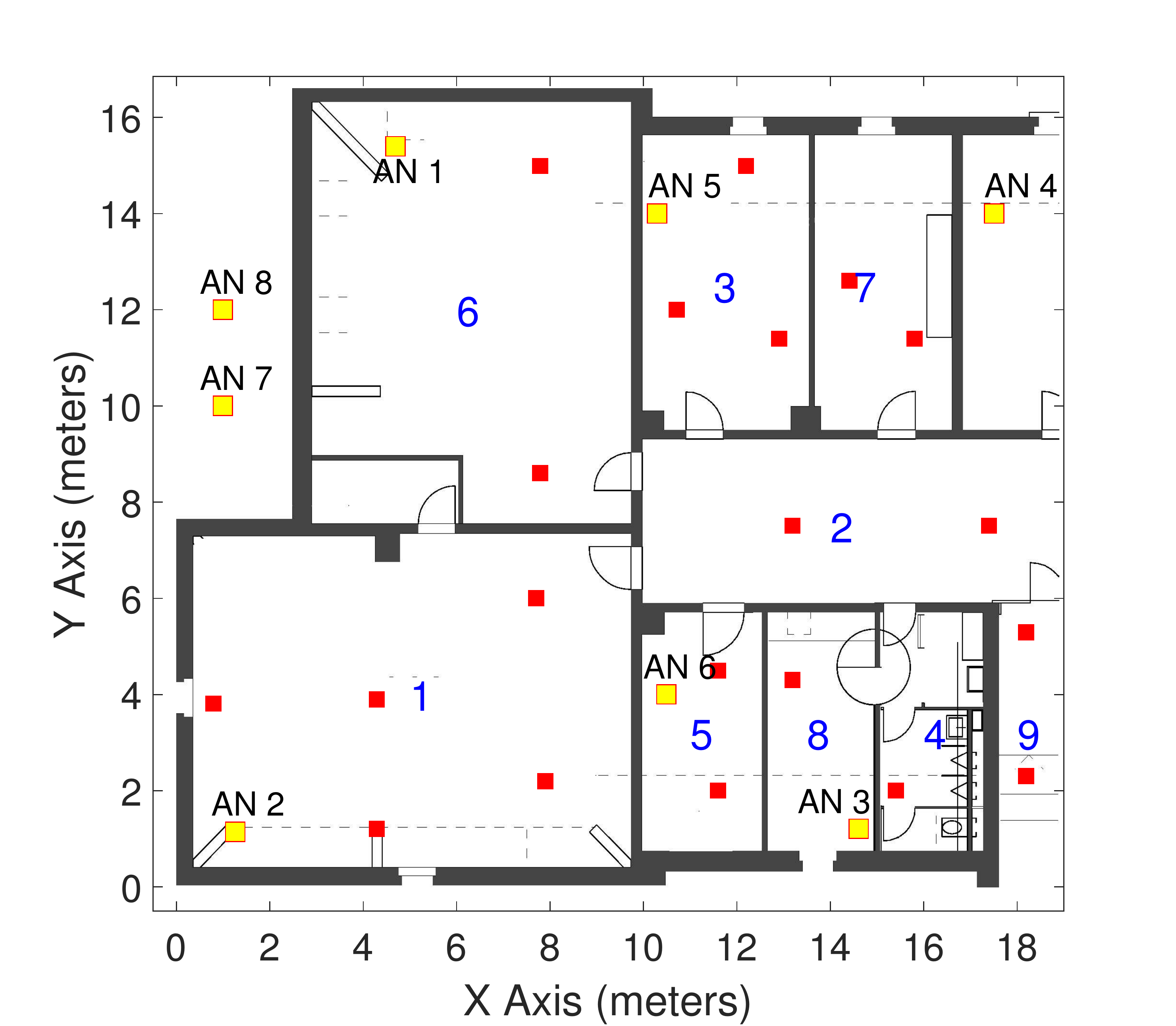} \label{scenario2}}
	\caption{Anchor Nodes distribution and room definition (Square Points: Anchor Nodes; blue numbers: room identifier). 5 ANs are considered in scenario 1, whereas 8 ANs are deployed in scenario 2.}
	\label{scenariosAnDistribtion}
\end{figure*}

Considering that the landmark detection accuracy can be influenced by some parameters of the environment, we conduct some experiments to determine how environmental parameters such as position of AP or number of APs influence the accuracy of the Wi-Fi-based fingerprinting landmark detection approach. Additionally, we perform experiments to show how the accuracy is improved by integrating magnetic field (MF) values and Wi-Fi readings to establish the fingerprinting data. As shown in Figure \ref{scenariosAnDistribtion}, we define 9 wall separated areas in our environment. Hereafter, we refer to these areas as rooms. In order to study the effect of different environmental configurations, we define two scenarios by varying some environmental parameters: the first scenario contains 5 ANs, the second scenario includes 8 ANs. In scenario 1 the positions of the 5 ANs are known, whereas in scenario 2 the positions of AN7 and AN8 are unknown. Figure \ref{scenariosAnDistribtion} presents the physical distribution of ANs in both scenarios. 

Internal parameters of learning-based algorithms are optimized from training data. Additionally, certain algorithms also have parameters that are not optimized during the training process. These parameters are called hyperparameters of the learning-based algorithm. Due to hyperparameters have significant impact on the performance of the learning-based algorithm, we use a nested cross validation technique to adjust them \cite{MLTechniques}. The nested cross validation technique defines an inner and outer cross validation. The inner cross validation is intended to select the model with optimized hyperparameters, whereas outer cross validation is used to obtain an estimation of the generalization error. Ten-fold cross validation was applied on both inner and outer cross validation. The classifiers were optimized over a key of hyperparameters. For KStar, we optimized  global blend percentage ratio hyperparameter, kernel type function, $c$ and $\gamma$ for SVM, the confidence factor for J48, number of hidden layers and neurons per layer for MLP. Based on the parameter optimization process, we established the optimal hyperparameter values for the classifiers as follows: global percent ratio of 30\% for KStar, single order polynomial kernel, $c=1$, $\gamma=0.0$ for SVM, single hidden layer with 10 neurons for MLP, and confidence factor of 0.25 for J48.

To test the room landmark detection performance, we performed three experiments. 
The first experiment tests room recognition performance in scenario 1. For this experiment, the landmark fingerprinting database for landmark detection contains 3712 instances, which were collected from the whole indoor environment. Additionally, we test two approaches regarding the structure of the fingerprint: Wi-Fi readings alone and Wi-Fi with magnetic field. The second experiment tests the influence of the position of the ANs over the room recognition accuracy. In this experiment we use the same number of ANs (AN1,AN2,AN3,AN4,AN5) as in experiment 1. Positions of AN3 and AN4 change with respect to scenario 1. Distance between AN4 and AN5 is increased. AN3 is placed on the right down corner of room 8. Figure \ref{scenario2} shows the physical positions of the ANs. The landmark fingerprint database contains 3700 instances, which were collected over the whole environment. The third experiment is intended to test the influence of the number of Wi-Fi ANs that are used to define the fingerprint instance. This experiment was performed in scenario 2. The results of the landmark detection experiments can be found in subsection \ref{subsection:result_landmark}.

To build the landmark fingerprint database  we ask a person to walk randomly around each room holding the phone in the hand. Instances must be collected equally distributed along the whole area in each room. Instance collection rate is only constrained by computational capabilities of the WiFi sensor of the smartphone. Thus, in our experiments each instance was collected at a rate of 3 instances/second. Because our approach doest not need to predefine any survey point, the time needed to build the landmark fingerprint database is proportional to the number of collected instances multiplied by the instance collection rate. Table \ref{FingerprintDatabases} shows the time required to build the landmark fingerprint database for the landmark detection method.

\subsection{Ranging Experiment Setups}

The first step in ranging estimation is to take the initial measurements, which are aimed to train the environmental parameters $\alpha$ and $\beta$ required for the NLR model. Similar to our previous work \cite{ARealTimeIndoorTracking}, we estimate $\alpha$ and $\beta$ values based on several stationary points. These reference points are spread over the whole floor plan. Please find details about the NLR ranging method in our previous work \cite{ARealTimeIndoorTracking}.

To analyze the performance of LDPL and NLR over our indoor environment, we define 40 test points over the whole scenario 2. Ranges between AN6 and the 40 test points are calculated with NLR, LDPL and PFML ranging models. The difference between ranges reported by the propagation model and ground truth ranges are considered as the ranging error. 

\subsection{Localization Experiment Setups}
In order to determine the localization errors,  $20$ testing positions were defined in the interested area. The MT was held by a person who was asked to stand still at each testing point during the localization process. The Euclidean distance between positions reported by the localization system and ground truth positions is considered as localization error. Red points in Figure \ref{scenario2} represent the evaluated  positions for experiments.

All the system parameters were determined beforehand. Table \ref{RangingModelParameters} shows the environmental parameters $\alpha$,  $\beta$ defined for the NLR model and $\gamma$, $P(r_0)$ parameters defined for the LDPL model. Parameter $N^{'}$ of the sampling function is set to 10\%. The experiment includes two phases: an off-line learning phase, and an on-line query phase. During the learning phase, a person holding the smartphone continuously walks through the indoor areas and collects the fingerprints and other correlated data. As mentioned before, our approach only requires room landmark detection. Thereby, the off-line learning efforts and implementation complexity are significantly reduced. During the on-line query phase, a person holds the smartphone and walks through the areas and her/his locations are presented on the smartphone in real-time.

\begin{table}
	\centering
	\caption{Environmental parameters for NLR ranging model}
	\begin{tabular}{|l|l|l|l|l|l|} \hline
	\textbf{AN}&\textbf{$\alpha$}&\textbf{$\beta$}&\textbf{$\gamma$}&\textbf{$P(r_0)$}\\ \hline
 AN1&1.264&-0.03614&2.7&-30\\
 \hline
 AN2&1.278&-0.03711&2.7&-32\\
   \hline
 AN3&0.3701&-0.05153&2.7&-30\\
   \hline
 AN4&0.5663&-0.05153&2.7&-31\\
   \hline
 AN5&0.3496&-0.05328&2.7&-30\\
   \hline
 AN6&0.9739&-0.03578&2.7&-30\\
   \hline
 \end{tabular}
	\label{RangingModelParameters}
\end{table}

\section{Experiment Results}
\label{Evaluation}
In this section, we discuss and analyze the results of each individual subcomponents, and the integrated real-time localization accuracy.

\subsection{Landmark Detection Accuracy}
\label{subsection:result_landmark}
This section discusses the accuracy of the landmark detection model when different classifiers are used. When comparing the performance of classifiers, it is impossible to define a single metric that provides a fair comparison in all possible applications.  Therefore, several measurements have been proposed in the literature \cite{ClassifierComparison}. We focus on the metrics of prediction accuracy and model building time. Accuracy refers to the percentages of correct room detection, and the model building time refers to the time that the algorithm requires to build the classification model.

\begin{figure}
	\centering
	\includegraphics[scale=0.28]{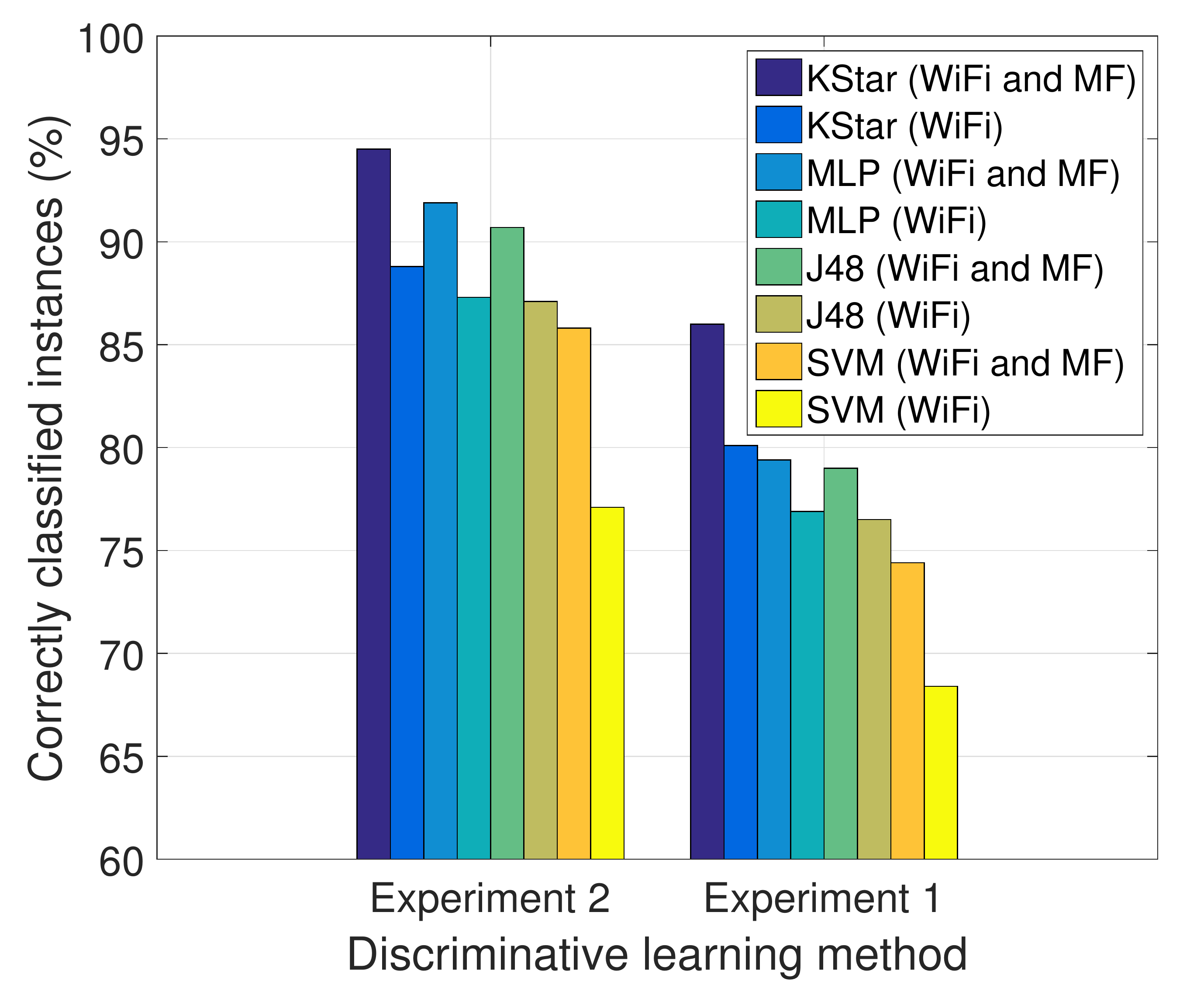}
	\caption{Landmark prediction performance in Experiment 1 and Experiment 2.}
	\label{DiscriminativePerformance}
\end{figure}



Figure \ref{DiscriminativePerformance} shows performance validation of the selected learning-based classifiers obtained from Experiment 1. The best performance is reached by the KStar classifier, which achieves 86\% of instances correctly classified when the fingerprint is composed by Wi-Fi and MF readings. By using Wi-Fi and MF readings in the room recognition process, the accuracy is improved in all tested classifiers. The maximum improvement takes place on SVM with 6\%. However, the KStar classifier achieves the best overall performance with 86\% of instances correctly classified. Figure \ref{DiscriminativePerformance} shows the evaluation results of the classifiers obtained from experiment 2. Percentage of correctly classified instances  is increased for all tested classifiers. This is due to the fact that the similarity of signatures in room 3 and room 7 decreases when the distance between AN4  and A5 increases. Therefore, this experiment shows that the position of the ANs influences room recognition accuracy. The KStar classifier achieves the best performance with 94.5\% of instances correctly classified. Table \ref{TClassNumberAN} presents the classifier validation results obtained from Experiment 3. KStar achieves the best performance, its percentage of instances correctly classified increases in 2.8\% by increasing the number of ANs from 5 to 8. However, the improvement is just 0.1\% from 7 to 8 ANs. Therefore, increasing the number of ANs improves the accuracy of the classifier. Nevertheless, after a certain number of APs is used, the improvement of adding more APs is almost negligible in all tested classifiers.

\begin{table}[h]
	\centering
	\caption{Classification performance vs number of ANs, Experiment 3 (scenario 2).}
	\begin{tabular}{|c|c|c|c|}\hline
		\textbf{ANs}
        &\textbf{Classifier}
		&\textbf{Accuracy (\%)}
        &\textbf{Training time (s)}
		 \\ \hline
		\multirow{4}{*}{6 ANs}&KStar&96.2&0\\ \cline{2-4}
		&MLP&94.0&6.88\\ \cline{2-4}
        &J48&92.3 &0.04\\ \cline{2-4}
        &SVM& 91.6&3.16\\ \hline   
        \multirow{4}{*}{7 ANs}&KStar&97.2&0\\ \cline{2-4}
		&MLP&95.3&8.17\\ \cline{2-4}
        &J48&93.1 &0.15\\ \cline{2-4}
        &SVM&94.4&12.87\\ \hline       
       \multirow{4}{*}{8 ANs}&KStar&97.3&0\\\cline{2-4}
		&MLP&95.8&8.53 \\ \cline{2-4}
        &J48&92.3 &0.23\\ \cline{2-4}
        &SVM&95.5&12.93\\ \hline     
	\end{tabular}
	\label{TClassNumberAN}
\end{table}

In indoor environments, measured values of RSSI vary according to locations. However, it is expected that these values will remain similar on nearby positions. For example, on locations close to landmark borders, high similarities will be observed on RSSI values. These similarities could lead to misclassification problems. The KStar classifier outperforms J48, SVM and MLP in terms of accuracy and building time on the three experiments. 
This is because KStar is an instance-based learner algorithm, which uses entropy as a distance measure to determine how similar two instances are. Distance between instances is defined as the complexity of transforming one instance into another. Thus, this method is more sensitive to slight variations upon the instance as unity. Unlike KStar, J48 builds the classification model by parsing the entropy of information at attribute level. It means that J48 measures entropy in the attribute domain to decide which attribute goes in a decision node. Therefore, the classification model is prone to misclassification in this specific landmark detection problem. 
Although SVM shows good performance, optimized hyperparameter values affect the smoothness of the decision boundary. It could lead to a certain degree of overfitting on the model. Moreover, building time on SVM is the highest compared to other classifiers.
MLP, by means of sigmoid function, achieves about 95.8\% of accuracy when using 8 ANs. However, its building time is 99.99\% and 99.98\% higher than KStar and J48 respectively. Therefore,  as shown on Table \ref{TClassNumberAN}, it is clear that the KStar classifier achieves better performance than J48, SVM and MLP for the room landmark detection problem.

\subsection{Ranging Accuracy}
\label{subsection:result_ranging}

\begin{figure}
	\centering
	\includegraphics[scale=0.25]{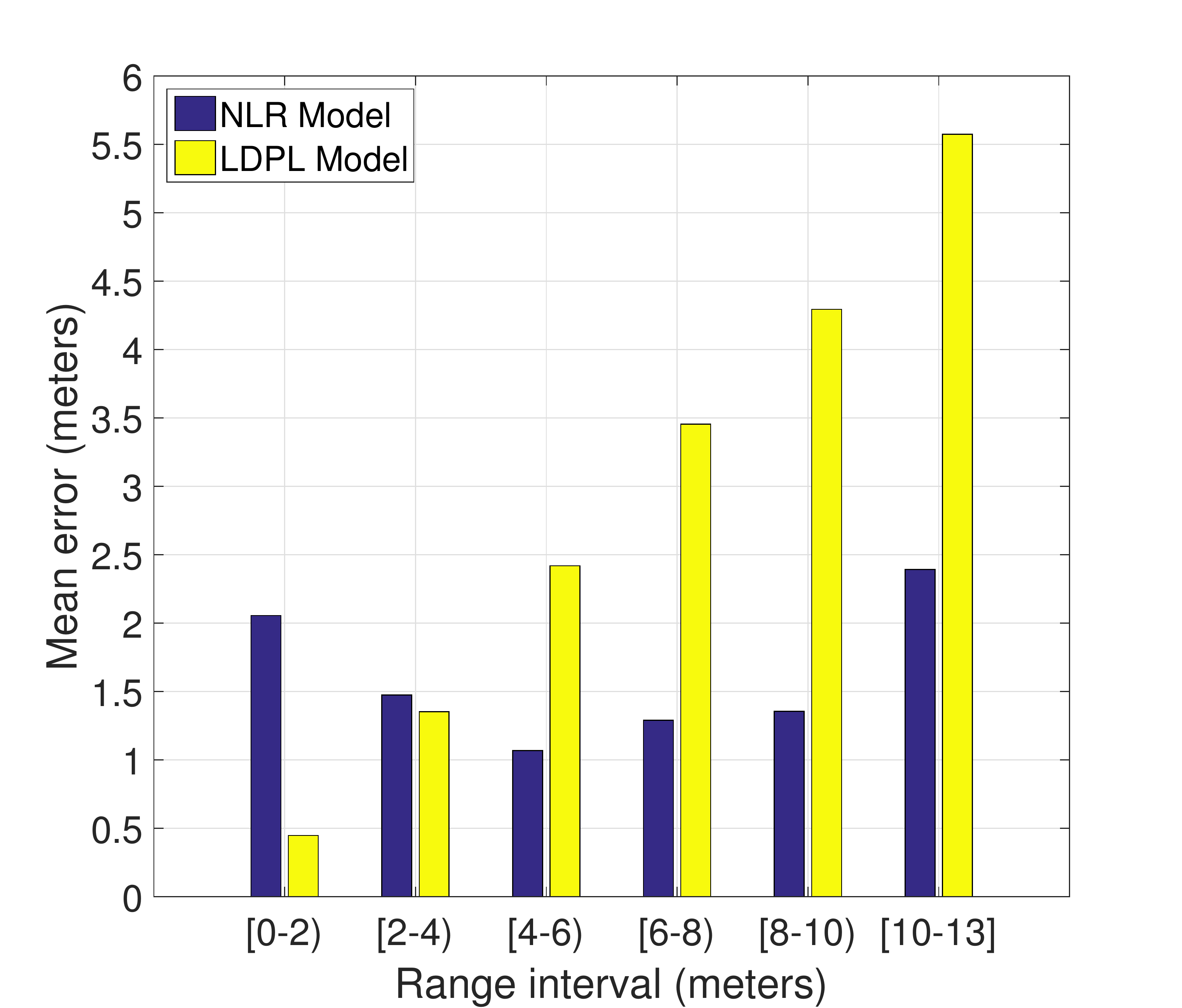}
	\caption{Propagation Models Error in Range intervals. LDPL perform better in short distances, whereas NLR perform better in large distances.}
	\label{rangingError}
\end{figure}

\begin{figure}
	\centering
	\includegraphics[scale=0.25]{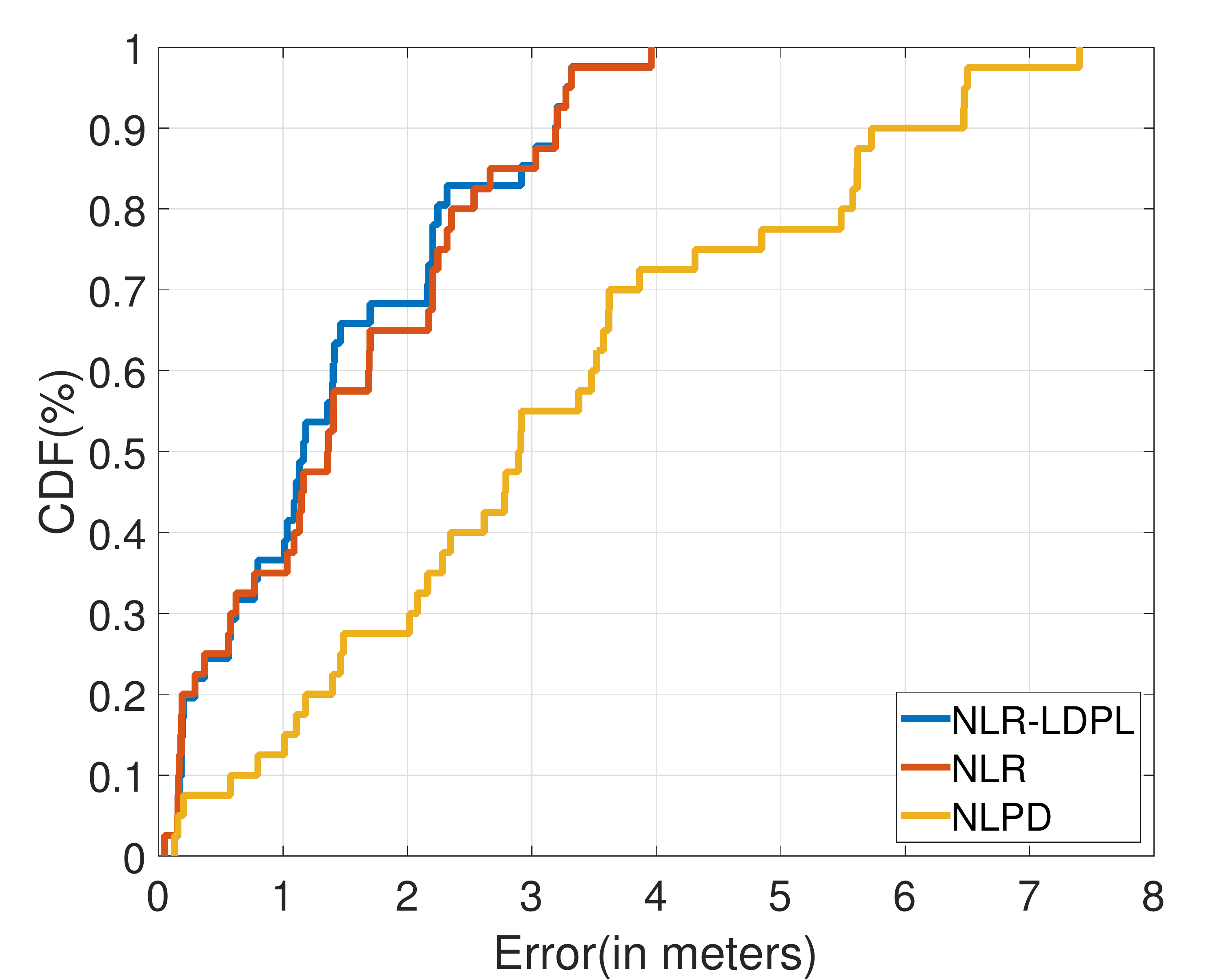}
	\caption{CDF ranging methods}
	\label{CDFrangingError}
    \vspace{-5pt}
\end{figure}

This section discusses the results of the proposed ranging model. Figure \ref{rangingError} shows the mean error of NLR and LDPL models considering range intervals. Table \ref{TRangingPropagationM} summarizes the ranging mean error of NLR, LDPL and NLR-LDPL models. Figure \ref{CDFrangingError} depicts the CDF of ranging errors for the signal propagation models. Considering the 40 test points, NLR achieves $1.48m$ of mean error, whereas LDPL achieves $1.79m.$ Therefore, NLR accomplishes better performance than LDPL. However, it is worth to notice that LDPL performs better than NLR when the range is shorter than $5m$, meaning that LDPL overcomes NLR in LOS conditions. In this case, LDPL achieves $1.01m$ of mean ranging error, whereas the mean error of NLR is $1.79m$. Thereby, LDPL achieves better performance for short distances, and NLR achieves better performance for large distances. This is because parameters of the LDPL model are defined in LOS conditions, whereas parameters of NLR are defined considering LOS and NLOS conditions. As shown in Table\ref{TRangingPropagationM}, our proposed NLR-LDPL ranging model outperforms NLR and LDPL. This is because NLR-LDPL considers LOS and NLOS conditions to apply the most suitable ranging method between NLR and LDPL. 

\begin{table}[h]
	\centering
	\caption{NLR and LDPL ranging errors.}
	\begin{tabular}{|c|c|c|}\hline
		\textbf{Model}
        &\textbf{Error(all points)}
		&\textbf{Error($<$5m points)}
		 \\ \hline	
		NLR&1.48 m.&1.79 m.\\ \hline
		LDPL&3.16 m.&1.01 m.\\ \hline
        NLR-LDPL&1.31 m.&1.01 m.\\ \hline
	\end{tabular}
	\label{TRangingPropagationM}
\end{table}

\subsection{Localization Accuracy}
\label{subsection:result_localization}

In this section we focus on the performance of the whole localization approach. 
A performance comparison between our approach and state of the art fingerprinting-based and landmark-based approaches would be biased. This is because most of those approaches are environmentally dependent (i.e., they rely on the presence of numerous landmarks in the environment). Therefore, their performance will vary on different environment conditions. Moreover, unlike aforementioned approaches, our localization solution only requires room landmark detection. However, to show the performance superiority of our solution, we compare the performance of our method with the Nonlinear Least Squares trilateration (NLST) algorithm,  K-nearest neighbors (KNN) fingerprinting and Kalman Filter-based (KF) localization methods. Hereafter, we refer to our particle filter machine learning localization approach as the PFML localization method.

For evaluation, we collect the metrics of CDF (Cumulative Distribution Function) of localization errors, mean tracking error, and standard deviation of localization errors.

\begin{figure*}[htp]
	\centering
	\subfloat[Confidence Interval]{\includegraphics[scale=0.25]{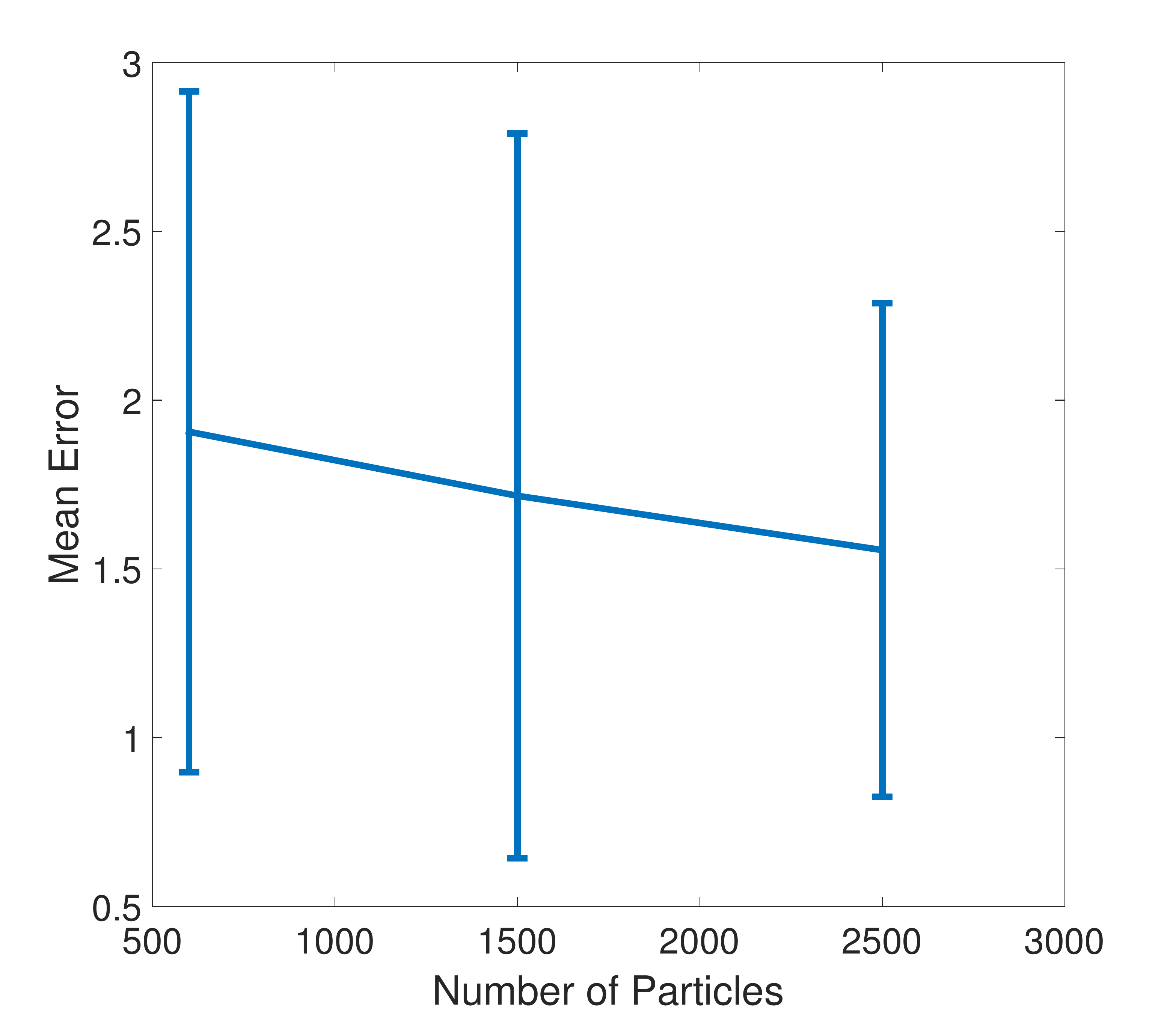}\label{confidenceInterval}}
	\subfloat[Localization error CDF]{\includegraphics[scale=0.25
    ]{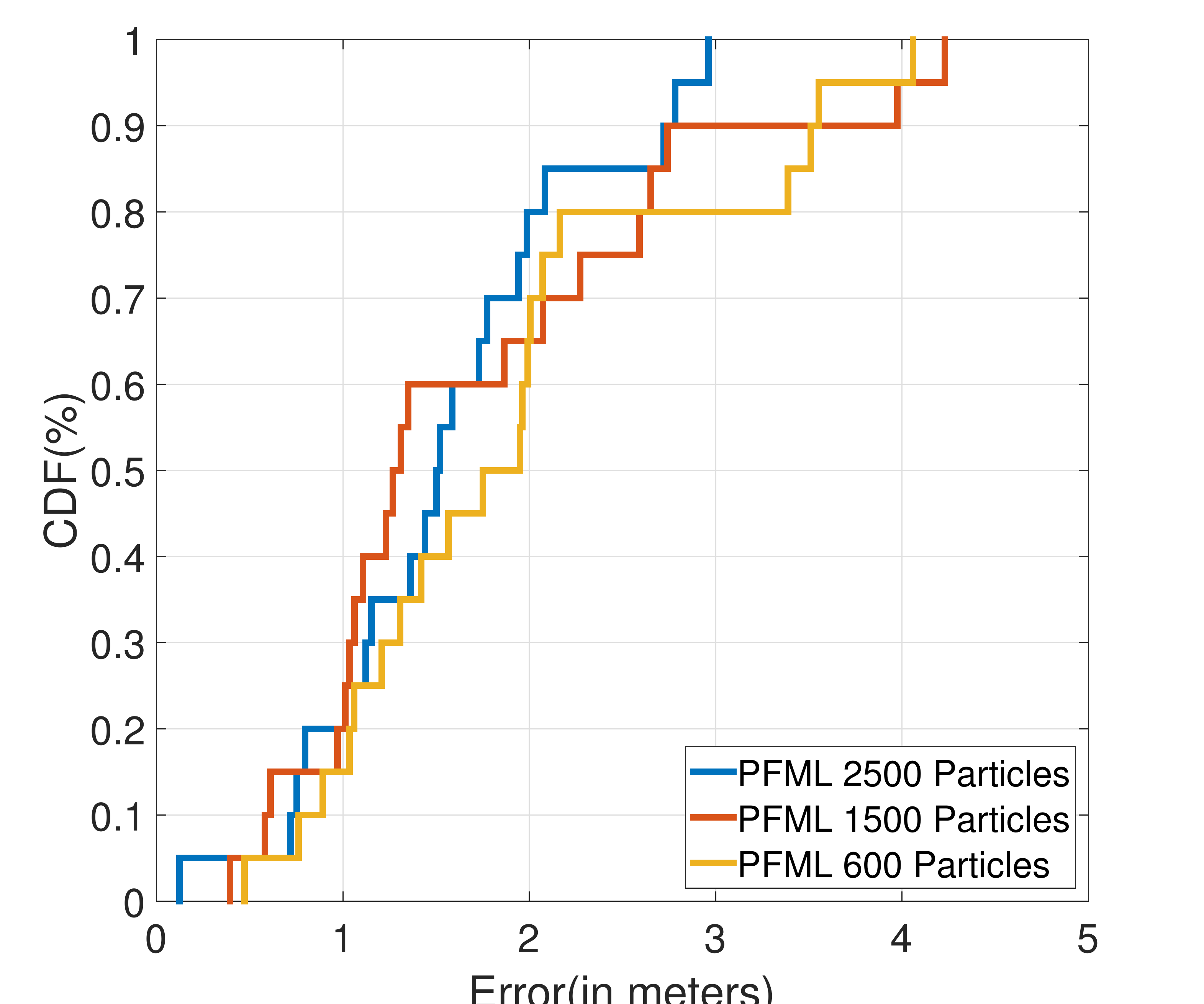} \label{LocalizationErrorParticle}}
	\caption{Localization performance vs number of particles }
\end{figure*}    

\begin{figure}
	\centering
	\includegraphics[scale=0.25]{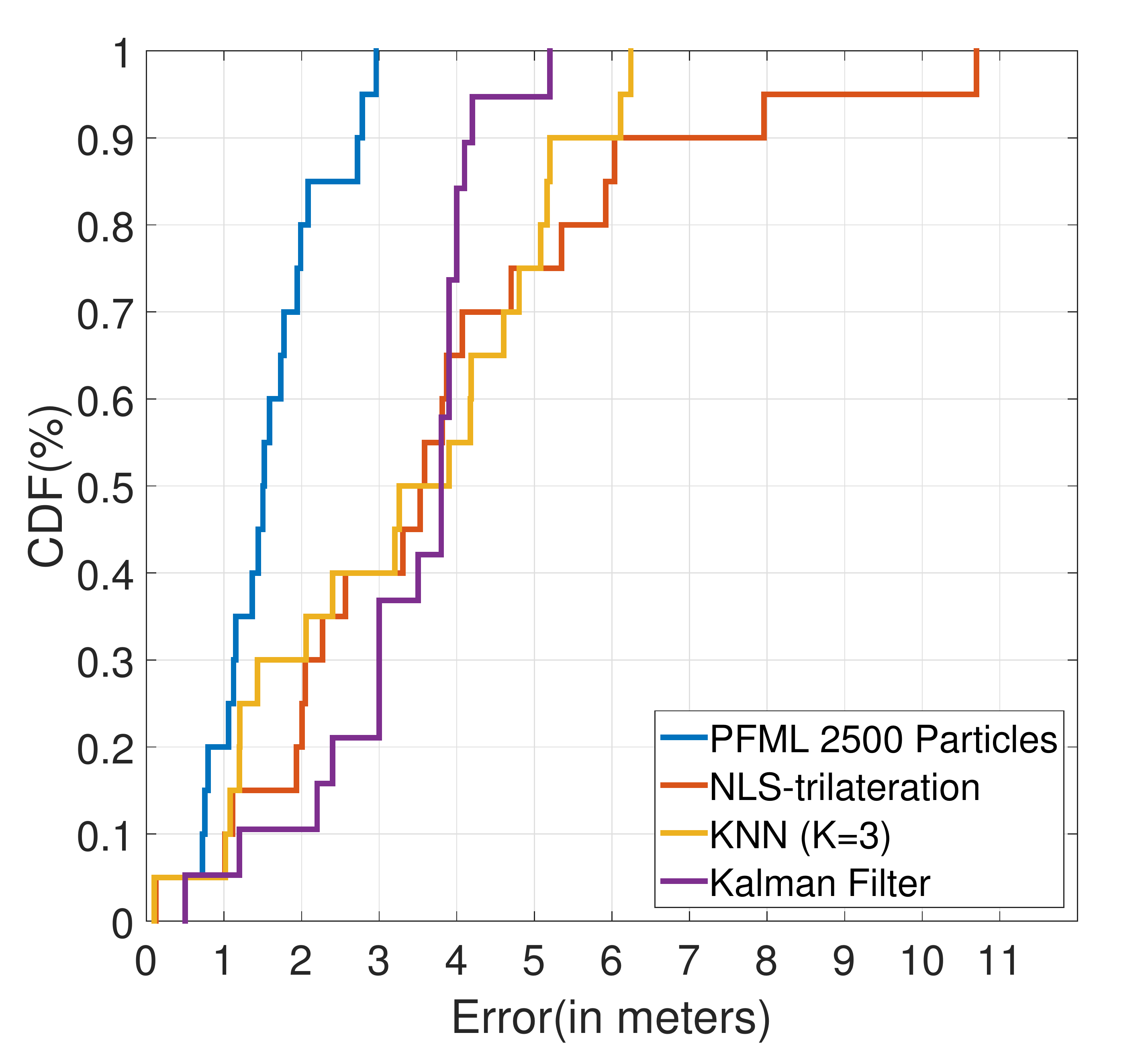}
	\caption{Localization error CDF PFML, NLST, KNN and KF}
	\label{LocalizationError}
\end{figure}
\begin{figure}
	\centering
	\includegraphics[scale=0.25]{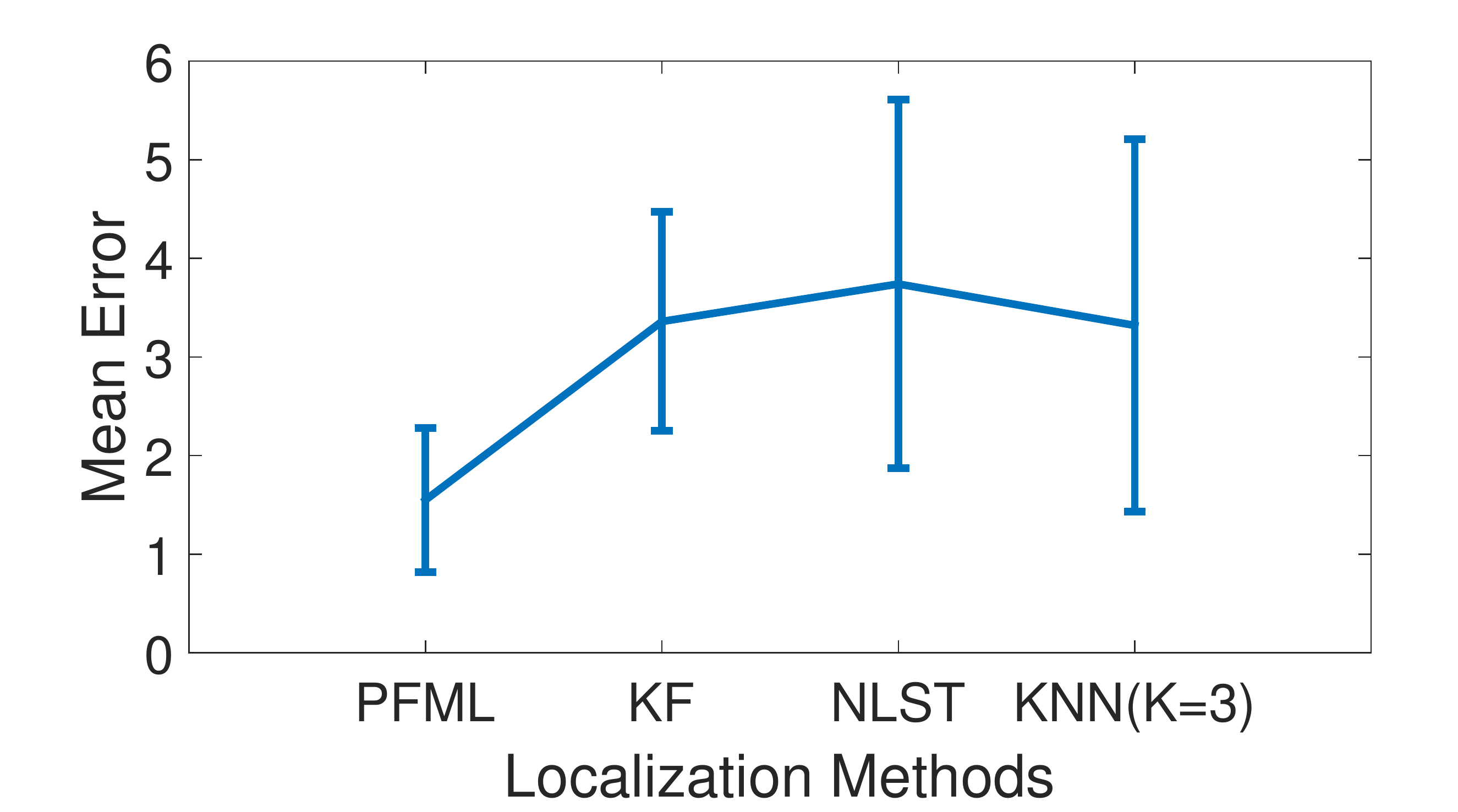}
	\caption{Confidence interval PFML,NLST,KNN}
	\label{confidenceIntervaLPF-NLS}
    \vspace{-5pt}
\end{figure}

\begin{figure*}
	\centering
	\subfloat[Localization error \textit{vs} number of ranging ANs]{\includegraphics[scale=0.25]{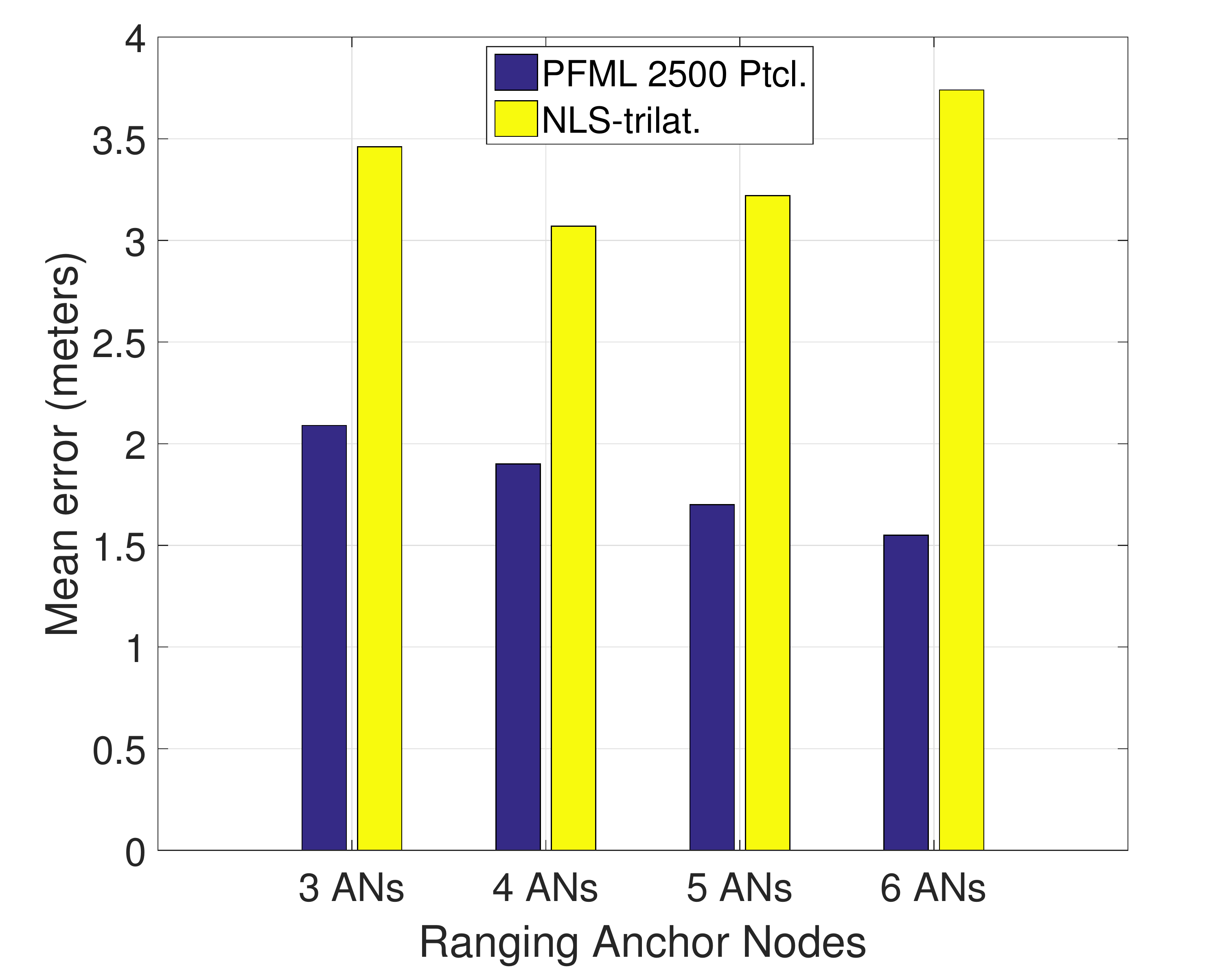}\label{barLocalizationPerformance}}
	\subfloat[PFML Localization error CDF \textit{vs} number of ranging ANs]{\includegraphics[scale=0.25
    ]{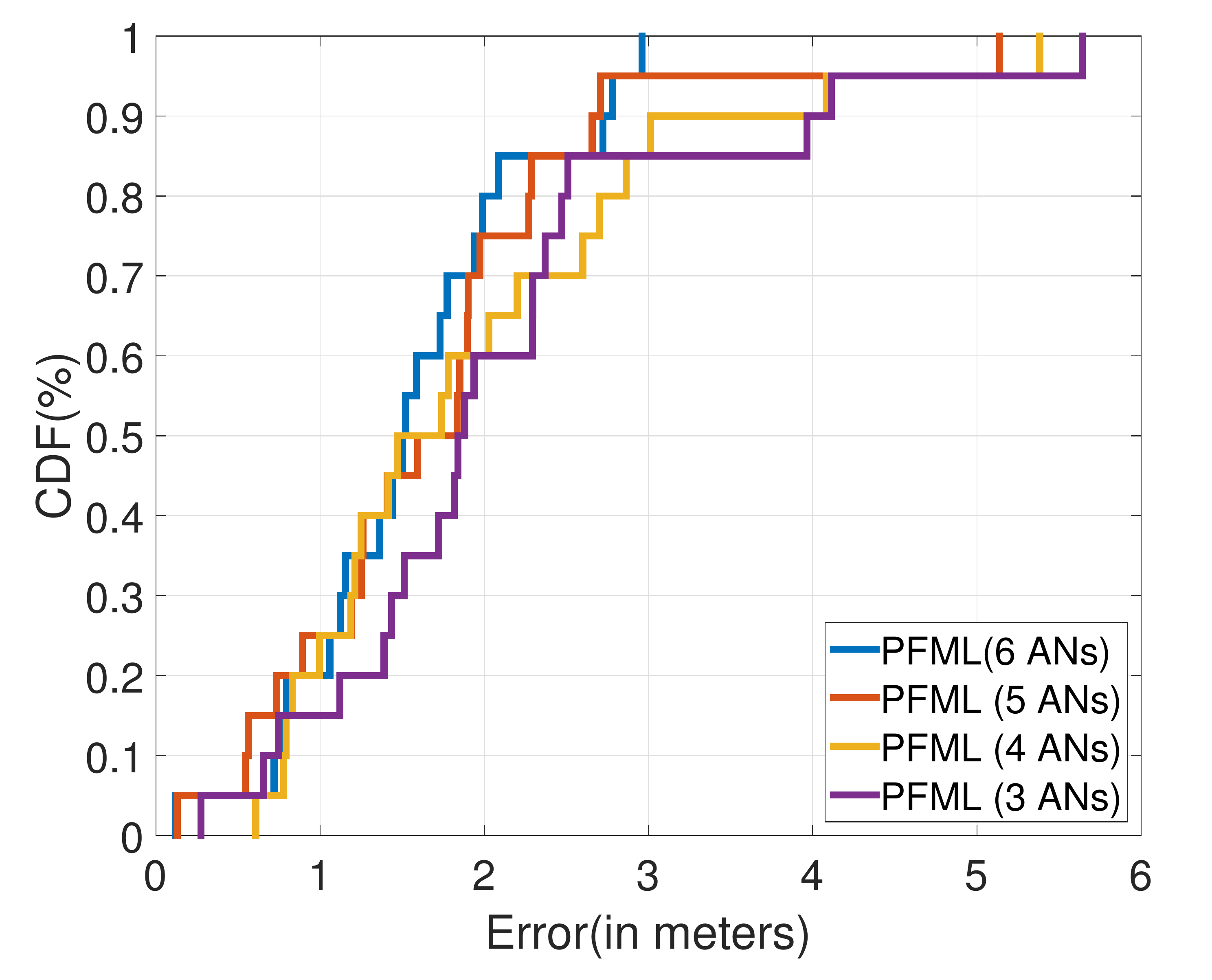} \label{LocalizationErrorANs}}
	\caption{Localization performance \textit{vs} number of ranging ANs}
\end{figure*}

\subsubsection{Performance vs Number of Particles}
This set of experiments was conducted in scenario 2 with 6 ANs for ranging estimation and 8 ANs for landmark recognition. Further details about the environment configuration can be seen in Figure \ref{scenario2}. The number of particles are 600, 1500, 2500. Figure \ref{confidenceInterval} depicts the confidence intervals resulting from this set of experiments. 
Figure \ref{LocalizationErrorParticle} shows the CDF of localization errors for PFML localization methods. Table \ref{tLocalizationError} summarizes the average of localization errors, standard deviation and $90\%$ accuracy. PFML achieves the minimum mean error of $1.55m$ with 2500 particles, which spend approximately $455$ milliseconds of processing time. It is worth to notice that this configuration also yields the lowest standard deviation: $0.73m$.
Our localization approach achieves the best performance by using 2500 particles. However, there is not a big gap among the performance values in all tested particle numbers in this experiment. This is because PFML defines a discrete system space to represent the environment. Thus, most of the potential system states from the posterior distribution can be covered by small number of particles. More particles will significantly increase the computation overhead, which degrades the overall system performance. \\
\subsubsection{Performance vs Number of Ranging ANs}
In this experiment we validate the robustness of our localization approach by varying the number of ranging ANs. We select the number of particles that yield the best performance of PFML. Therefore, we setup the PFML method to 2500 particles. Moreover, we compare the performance of the PFML approach to NLST. Figure \ref{barLocalizationPerformance} depicts the mean localization error achieved  by PFML and NLST methods. Figure \ref{LocalizationErrorANs} shows the CDF of localization errors of our proposed localization approach. Table \ref{tLocalizationErrorPFML} summarizes the mean localization error, standard deviation and 90\% accuracy of the PFML method. Table \ref{tLocalizationErrorNLS} shows the mean localization error and standard deviation of NLST.

\begin{table}
	\centering
	\caption{PFML Performance vs Number of ranging ANs}
	\begin{tabular}{|c|c|c|c|}\hline
		\textbf{Configuration}
		&\textbf{Mean error}
		&\textbf{S.D}
		&\textbf{90\% Acc.} \\ \hline
		PFML 6 ANs&1.55m&0.73m&2.8m\\ \hline
        PFML 5 ANs&1.7m&1.1m&2.7m\\ \hline
        PFML 4 ANs&1.95m&1.22m&3.0m\\ \hline
        PFML 3 ANs&2.09m&1.26m&2.5m\\ \hline
		NLST 6 ANs&3.05m&2.74m&7.1m \\ \hline
	\end{tabular}
	\label{tLocalizationErrorPFML}
\end{table}

\begin{table}
	\centering
	\caption{NLST Performance vs Number of ranging ANs}
	\begin{tabular}{|c|c|c|}\hline
		\textbf{Configuration}
		&\textbf{Mean error}
		&\textbf{S.D}
		\\ \hline
		NLST 6 ANs&3.05m&2.74m\\ \hline
        NLST 5 ANs&3.22m&2.48m\\ \hline
        NLST 4 ANs&3.07m&1.3m\\ \hline
        NLST 3 ANs&3.46m&1.42m\\ \hline
		
	\end{tabular}
	\label{tLocalizationErrorNLS}
\end{table}

Based on these results, we can highlight the following observations. First, PFML achieves more stable and higher accuracy than the NLST method in all tested cases. This is because unlike NLST, PFML determines the target localization from a Posterior Probability Distribution (PPD). This PPD is built after considering multiple information sources such as ranges and room landmark detection. Second, by using 5 ranging ANs, PFML outperforms NLST by around $47.2\%$ considering  mean error. The  standard deviation of PFML is $55.6\%$ smaller than for NLST. Third, by considering 4 ranging ANs, the standard deviation of our localization approach is $6.15\%$ smaller than for NLST. PFML performs better than NLST by around $36.5\%$ considering the mean error. Fourth, by considering 3 ranging ANs, PFML outperforms NLST by around $39.59\%$ and $11.2\%$ regarding mean error and standard deviation, respectively.

Experiment results show that the accuracy of PFML increases as more ranging ANs are used, whereas NLST accuracy does not significantly change.  This reflects that unlike NLST, our localization approach is able to exploit diverse environment information sources (e.g., magnetic filed and wireless signal propagation) to  increase localization accuracy.  \\

\subsubsection{Comparison with Other Systems}
\label{subsection:result_comparison}
Performance comparison with existing solutions can proof the superiority of the proposed system. However, for indoor localization, it is rather hard to implement all the specific details of an existing solution and repeat the identical experiment to get the same results that were collected in another physical indoor environment. Therefore, similar to many other indoor localization systems, we compare our system with classical indoor localization solutions, such as fingerprinting, Kalman filter-based (KF) and trilateration approaches.

Figure \ref{LocalizationError} depicts the CDF of localization error for the best performance of PFML, NLST and the KNN method (K=3), whereas Figure \ref{confidenceIntervaLPF-NLS} shows the confidence intervals of the best performance of PFML, KF, NLST and KNN (K=3) methods. 
As shown in Table \ref{tLocalizationError}, NLST achieves around $8.0m$ for $90\%$ accuracy, the mean error is $3.79m$ and the standard deviation is $2.52m$. For $K=3$, the KNN localization method achieves around $6.1m$ for $90\%$ accuracy, the mean error is $3.32m$ and the standard deviation is $1.89m$. The KF approach achieves around $4.1m$ for $90\%$ accuracy, the mean error is $3.36m$ and the standard deviation is $1.11m$.

\begin{table}
	\centering
	\caption{Performance vs Number of Particles}
	\begin{tabular}{|c|c|c|c|}\hline
		\textbf{Configuration}
		&\textbf{Mean error}
		&\textbf{S.D}
		&\textbf{90\% Acc.} \\ \hline
		PFML (2500 Ptc.)&1.55m&0.73m&2.8m\\ \hline
        PFML (1500 Ptc.)&1.71m&1.073m&2.62m\\ \hline
        PFML (600 Ptc.)&1.91m&1.01m&3.5m \\ \hline    
        NLST&3.79m&2.52m&8.0m \\ \hline
        KNN (N=1)&3.51m&2.1m&6.2m \\ \hline
        KNN (K=3)&3.32m&1.89m&6.1m \\ \hline
         Kalman Filter&3.36m&1.11m&4.1m \\ \hline
	\end{tabular}
	\label{tLocalizationError}
\end{table}
\begin{table}
	\centering
	\caption{Offline survey time of different approaches}
	\begin{tabular}{|c|c|c|}\hline
		\textbf{Localization method}
		&\textbf{Instances}
		&\textbf{Survey time} \\ \hline
		PFML(landmark detection + ranging)&3712+20 &21+28=49 min\\ \hline
        KNN&5060&460 min\\ \hline 
        NLST (ranging)&20&28 min\\ \hline 
        LS-SVM-WA \cite{Generative+Discriminative}&8056 &403 min\\ \hline 
	\end{tabular}
	\label{FingerprintDatabases}
\end{table}
Our localization approach outperforms NLST by around $65\%$, KNN  by around $54.1\%$ and KF by around $31.7\%$ regarding $90\%$ accuracy. Moreover, confidence intervals of PFML are around $20\%$ smaller than for KF and $50\%$ smaller than for NLST and KNN. The mean error of the PFML approach is $53.8\%$, $59.1\%$ and $53.31\%$ better than for KF, NLST, and KNN respectively. Standard deviation is $71.03\%$ smaller than for NLST and $61.4\%$ smaller than for KNN. Moreover, it is worth to mention that because PFML requires only room landmark detection, the survey time of PFML is around $88\%$ smaller than the survey time employed to build the KNN fingerprinting database.

Experiment results show that because of the integration of machine learning-based room landmark detection, ranging models and particle filter methods, our localization method outperforms KF, NLST and KNN for accuracy and stability. Moreover, PFML can significantly reduce the off-line survey effort compared to common fingerprint-based and landmark-based localization approaches. Furthermore, although the ranging method for both PFML and NLST is the same,  PFML is able to achieve higher localization accuracy than NLST.

\subsection{Advantage of Using PFML Localization Method}
As mentioned before, PFML is able to reduce the off-line phase effort while still achieving high accuracy. Unlike traditional fingerprinting-based indoor localization methods, PFML does not define any fixed survey point to build the landmark fingerprint database. Thus, the offline phase becomes a simple process. The time required to build the landmark fingerprint database depends only on the number of collected instances and the instance collection rate given by the Wi-Fi sensor of the smartphone (i.e., 3Hz). Therefore, the time required to complete the offline phase in PFML is defined as follows:

\begin{equation}
	T_{offline,PFML}= I\cdot S_{rate}+ T_{ranging},
\end{equation}  

where $I$ is the number of collected instances, $S_{rate}$ is the instance collection rate.

The landmark fingerprint database of the PFML localization method contains $3712$ instances. Thus, building this database takes around $3712/(3\cdot 60)= 21$ minutes. Additionally, to estimate $\alpha$ and $\beta$ values for the ranging method, we define $20$ reference points. This process takes around $28$ minutes. Thus, the required time to complete the offline phase in PFML is around $21+28=49$  minutes.

The KNN radio map database contains $5060$ instances. We define the indoor area as a grid of square cells of $l=1m$. Each cell intersection defines a survey point. Thus, $85$ survey points were defined in the whole area of interest. Around $60$ instances were collected in each survey point. In order to compare KNN with our approach, we relate the required time to complete the offline phase in fingerprinting-based approaches to the number of survey points, the time needed to define each survey point, the time needed to go from one survey point to another during the site survey phase, number of collected instances and the instance collection rate.  Therefore, the total offline time in KNN is defined as follows:

\begin{equation}
	T_{offline, KNN}=Sp \cdot(t_{sp} + t_{sw})+ I\cdot S_{rate},
\end{equation}  

where $Sp$ is the number of survey points, $t_{sp}$  is the time needed to measure the coordinates of each survey point, $t_{sw}$ is the time needed to go from one survey point to another during the site survey phase, $I$  is the number of collected instances, $S_{rate}$ is the instance  collection rate. 

Empirically, we approximate $t_{sp}$ and $t_{sw}$ to $300$ and $5$ seconds respectively. Therefore, the required time to complete the offline phase in KNN is around $460$ minutes. Assuming that the effort is proportional to the time spent in the offline phase, PFML reduces the offline effort by around $89\%$ compared to fingerprint-based approaches. In LS-SVM-WA \cite{Generative+Discriminative}, 80 labeled and 72 unlabeled samples are collected on 53 fixed survey points, which takes $(53\cdot 305+ 8056)/60=403$ minutes. 

Table \ref{FingerprintDatabases} summarizes number of data instances that have been collected during the experiments and  survey time of PFML, KNN, NLST, and LS-SVM-WA \cite{Generative+Discriminative}. Clearly, PFML takes much less time and effort to deploy the system and perform the data measurement when compared to fingerprinting-based, landmark-based approaches and other works that focus on reducing the calibration efforts.

\section{Conclusions}
This work exploits an enhanced particle filter approach to fuse discriminative learning-based landmark detection, ranging methods and physical information of the environment to achieve high localization accuracy in complex indoor scenarios for smartphones. Since our approach requires only room landmark detection, we are able to significantly reduce the effort and time required in processing the off-line phase compared to traditional fingerprinting-based and landmark-based approaches. Moreover, to incorporate a suitable discriminative learning-based landmark detection method, we assess the room recognition performance with the most popular discriminative learning algorithms. We conduct experiments to analyze the impacts of anchor nodes' position on the landmark detection accuracy. Additionally, based on ranging experiment results, we propose a composed NLR-LDPL ranging method. To evaluate our localization system, we conducted extensive experiments in a complex office-like environment. Experiment results show that our approach can achieve an average tracking error of $1.55m$ and $90\%$ accuracy is $2.8m$. Compared to fingerprinting-based and landmark-based approaches, our solution requires much less time in the offline process, while keeping the accuracy at a high level. Our solution is more accurate and stable than the commonly used Kalman filter, nonlinear least square trilateration and KNN-fingerprinting methods. Furthermore, our proposed approach enables real-time localization at the smartphone without assistance of any additional server.

\section*{Acknowledgment}
This work was partly supported by the Swiss National Science Foundation via the SwissSenseSynergy project under grant number 154458.

\ifCLASSOPTIONcaptionsoff
  \newpage
\fi



%

%

\begin{IEEEbiography}[{\includegraphics[width=1in,height=1.25in,clip]{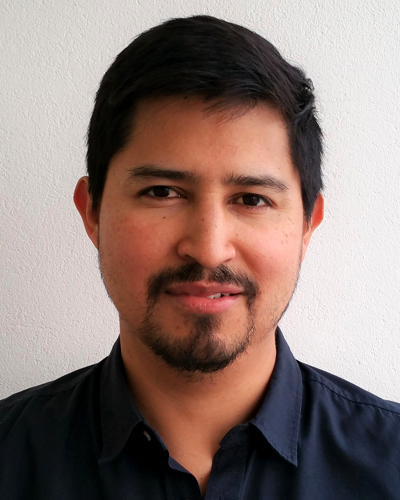}}]{Jos\'e Luis. Carrera V.}
received the B.S.E. degree from the National Polytechnic School from Equator and the MSc. degree in computer sciences from the Swiss Joint Master of Science in Computer Science program of the universities of Neuch\^atel, Fribourg and Bern in 2015. He is currently working toward the Ph.D. degree in the Institute of Computer Sciences of the  University of Bern. His research interests include Artificial Intelligence, Machine Learning, indoor localization and  distributed systems. 
\end{IEEEbiography}

\begin{IEEEbiography}[{\includegraphics[width=1in,height=1.25in,clip]{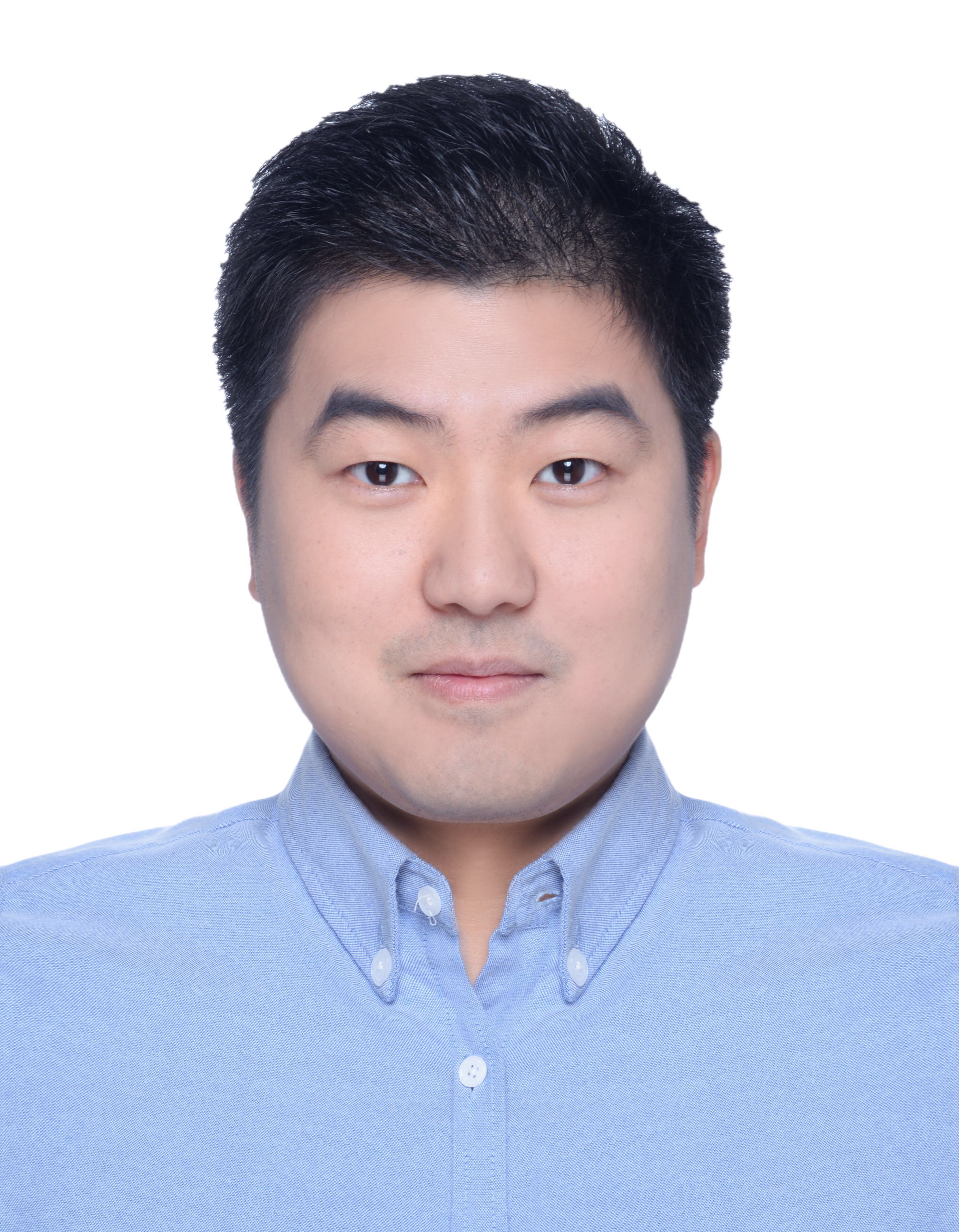}}]{Zhongliang Zhao}
received the Ph.D. degree from the University of Bern in 2014. He was a Senior Researcher with the University of Bern since then. He was multiple Work Package leaders in the EU FP7 project Mobile Cloud Networking, a Co-PI of the Sino-Swiss Science and Technology Cooperation project M3WSN. He is currently the Technical Coordinator of the Swiss National Science Foundation project SwissSenseSynergy, and Orange-funded Context Awareness Engine industry project.
\end{IEEEbiography}

\begin{IEEEbiography}[{\includegraphics[width=1in,height=1.25in,clip]{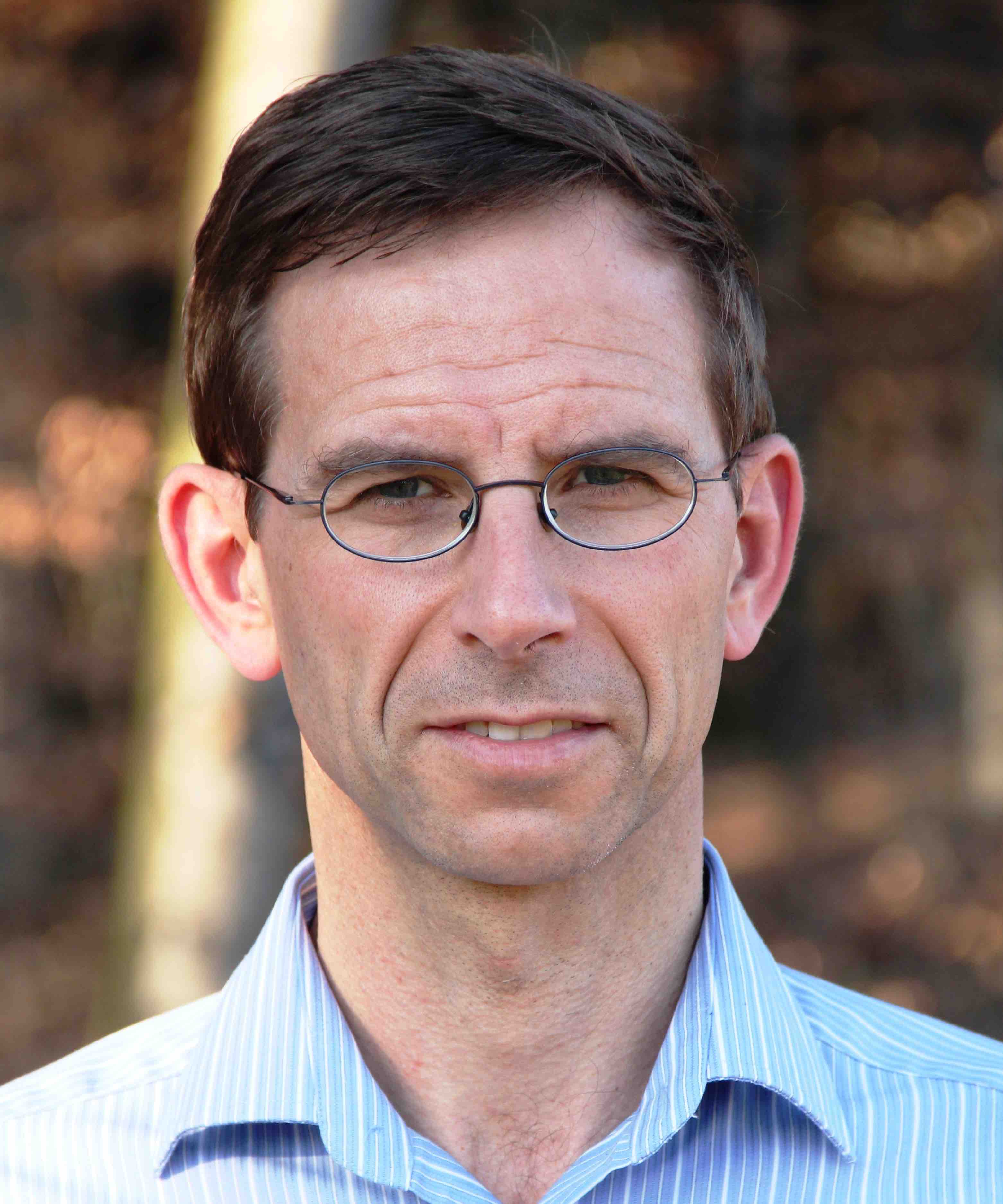}}]{Torsten Braun}
received the Ph.D. degree from the University of Karlsruhe, Germany, in 1993. Since 1998, he has been a Full Professor in computer science with the University of Bern. He has been the Vice President of the SWITCH Foundation since 2011. He received the Best Paper Award from the IEEE LCN 2001, WWIC 2007, EE-LSDS 2013, IFIP WMNC 2014, ARMSCC 2014 Workshop, and the GI-KuVS Communications Software Award in 2009.
\end{IEEEbiography}

\begin{IEEEbiography}[{\includegraphics[width=0.95in,height=1.25in,clip]{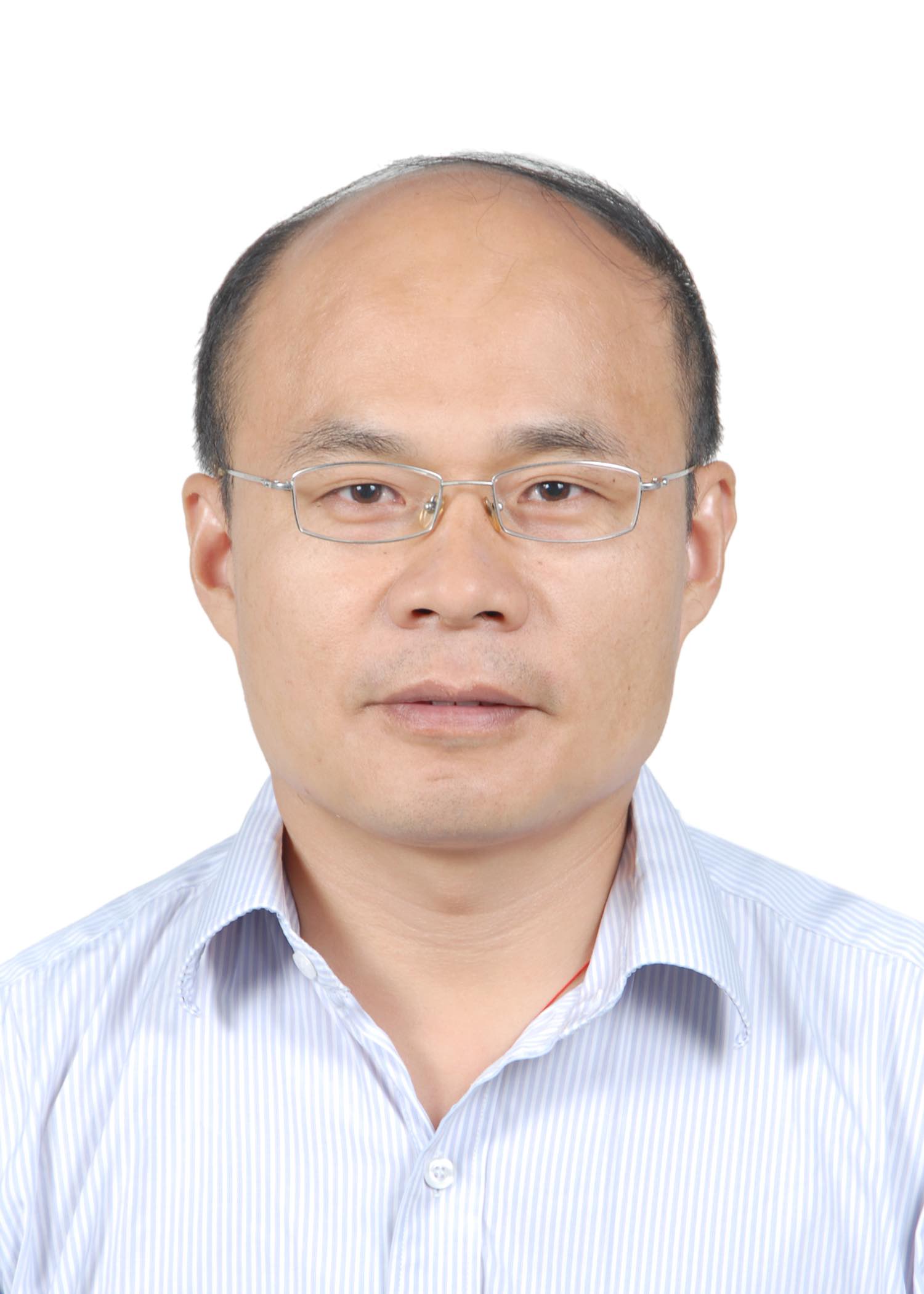}}]{Haiyong Luo} received the B.S. degree from Huazhong University of Science and Technology, Wuhan, China, in 1989, M.S. degree from the Beijing University of Posts and Telecommunications, Beijing, China, in 2002, and Ph.D. degree from the University of Chinese Academy of Sciences (CAS), Beijing, China, in 2008. He is currently an Associate Professor at the Institute of Computing Technology, CAS. His main research interests include pervasive computing, mobile computing, and the Internet of Things.
\end{IEEEbiography}

\begin{IEEEbiography}[{\includegraphics[width=0.95in,height=1.25in,clip]{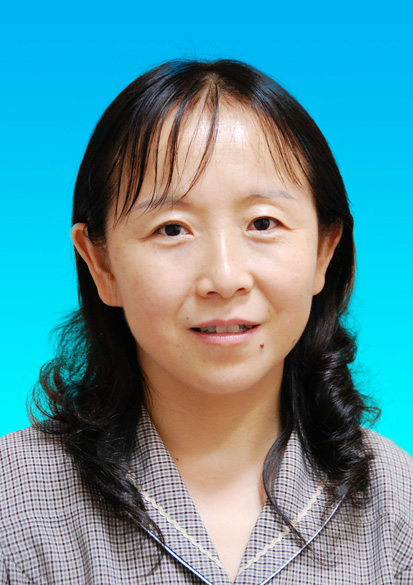}}]{Fang Zhao}
received the B.S. degree from Huazhong University of Science and Technology, Wuhan, China, in 1990, M.S. and Ph.D. degrees from Beijing University of Posts and Telecommunications (BUPT), Beijing, China, in 2004 and 2009. She is currently a Professor in School of Software Engineering, BUPT. Her current research interests include mobile computing, location-based services, and computer network.
\end{IEEEbiography}




\end{document}